\documentstyle[psfig]{l-aa}
\newcommand{\be}{\begin{equation}}
\newcommand{\ee}{\end{equation}}
\newcommand{\dd}{\mbox{d}}

\begin{document}
\thesaurus{12(02.01.2;02.13.2;08.06.2;09.10.1;11.14.1;11.10.1)}

\title{Magnetically-driven jets from Keplerian accretion discs} 
 
\author{Jonathan Ferreira}

\institute{Landessternwarte, K\"onigstuhl, D-69117 Heidelberg, Germany \\
jferreir@lsw.uni-heidelberg.de}

\date{Received March 30; accepted July 2, 1996}

\maketitle

\begin{abstract}

Non-relativistic, magnetically-driven jets are constructed by taking
self-consistently into account the feedback on the underlying 
accretion disc. It is shown that such jets are mostly described by the
ejection index $\xi= \dd \ln \dot M_a/\dd \ln r$, which is a local measure of
the disc ejection efficiency. This parameter is found to lie in a very
narrow range, due to constraints imposed by both the disc vertical
equilibrium and the steady transfer of angular momentum.

The investigation of global disc-jets solutions provided two important
results. First, it shows that the disc vertical equilibrium imposes a 
minimum mass flux ejected. Thus, one cannot construct jet models with
arbitrarily small mass loads. Second, their asymptotic behaviour critically
depends on a fastness parameter $\omega_A= \Omega_* r_A/V_{Ap,A}$, ratio of
the field lines rotation velocity to the poloidal Alfv\'en velocity at the
Alfv\'en surface. This parameter must be bigger than, but of the order of,
unity. 

Self-similar jets from Keplerian discs, after widening up to a maximum
radius whose value increases with $\omega_A$, always recollimate towards
the jet axis, until the fast-magnetosonic critical point is reached. It is
doubtful that such solutions could steadily cross this last point, the jet
either ending there or rebouncing. Recollimation takes place because of the
increasing effect of magnetic constriction. This systematic behaviour is
due to the large opening of the magnetic surfaces, leading to such an
efficient acceleration that matter always reaches its maximum poloidal
velocity. This ``over-widening'' stems from having the same ejection
efficiency $\xi$ in the whole jet.

Realistic jets, fed with ejection indices varying from one magnetic surface
to the other, would not undergo recollimation, allowing either cylindrical
or parabolic asymptotic collimation. The study of such jets requires full
2-D numerical simulations, with proper boundary conditions at the disc
surface.

\end{abstract}

\section{Introduction}

It is now widely accepted that accretion of matter onto a central object
plays a major role in astrophysics: in active galactic nuclei (AGN), young
stellar objects (YSOs), but also in evolved binary systems like cataclysmic
variables or low-mass X-ray binaries.

The standard theory of accretion discs (Shakura \& Sunyaev
\cite{shak:suny73}, Novikov \& Thorne \cite{novi:thor}, Lynden-Bell \&
Pringle \cite{lynd:prin}) assumes the presence of a turbulent viscosity
that allows the matter to loose its angular momentum and mechanical energy.
This energy is then radiated away at the disc surfaces, providing the
observed luminosity (e.g., Bregman \cite{breg}, Bertout et al.
\cite{bert88}). However, such a theory is unable to explain the origin,
acceleration and collimation of bipolar jets, that are emanating from radio
loud AGN and quasars (Bridle \& Perley \cite{brid:perl}), all YSOs (Lada
\cite{lada}) and some galactic objects like SS 433, Sco X-1 (Padman et
al. \cite{padm}) and ``microquasars'' (Mirabel et al. \cite{mira92}). 

In the vicinity of a black hole, the jet plasma has to come from a surrounding
accretion disc. Lynden-Bell (\cite{lynd}) suggested that in the framework
of a thick accretion disc, jets could be produced in the inner funnel
around the axis of the torus, and accelerated like in a De Laval
nozzle. Such a thick torus could be either radiation-supported (with a
super-Eddington accretion rate, e.g. Abramowicz et al. \cite{abra80}), or
ion-supported (with a sub-critical accretion rate, Rees et
al. \cite{rees82}). Jets would then be accelerated by radiation pressure
(Abramowicz \& Piran \cite{abra:pira}) or by the rotational energy of the
central black hole (Blandford \& Znajek \cite{blan:znaj}). This latter
process, unique possibility within the second scenario, invokes a large
scale magnetic field dragged in by the disc, braking the fastly rotating
hole and transferring its energy to matter (see also Phinney
\cite{phin83}). However, it has been shown that such thick tori are
violently unstable (Papaloizou \& Pringle \cite{papa:prin}, Zurek \& Benz
\cite{zure:benz}, Begelman et al. \cite{bege87}), casting therefore strong
doubts upon their viability to produce jets.  

The remaining scenario invokes a large scale magnetic field, anchored on a
geometrically thin (Keplerian) accretion disc (Blandford \& Payne
\cite{blan:payn}, hereafter BP82). These authors showed that this
magnetic field could brake the disc and carry away its angular
momentum. If jets carry a current, then the toroidal component of the
magnetic field would provide a tension that naturally confines the jet
(previously recognized by Chan \& Henriksen \cite{chan:henr}).

For protostars, the situation is more complex, since the observed jets could
be either stellar winds (Canto \cite{cant}, Hartmann et al. \cite{hart82},
Lago \cite{lago}) or disc winds (Pudritz \& Norman \cite{pudr:norm83},
Uchida \& Shibata \cite{uchi:shib}). However, both 
mass and momentum fluxes of the observed outflows are much higher than the
ones provided by the protostar luminosity, hence forbidding both thermally
and radiation pressure driven stellar winds (DeCampli \cite{deca}, K\"onigl
\cite{koni86}). The possibility remains that stellar magnetic fields play a
major role in producing a wind (Mestel \cite{mest}, Hartmann \& MacGregor
\cite{hart:mcgr}, Sakurai \cite{saku85}, Tsinganos \& Trussoni 
\cite{tsin:trus}, Sauty \& Tsinganos \cite{saut:tsin}). But the strongest
argument in favour of disc-driven jets is certainly the observed
correlation between signatures of accretion and ejection (Cabrit et
al. \cite{cabr90}, Hartigan et al. \cite{harti95}). 

To summary, one can conclude that both observational and theoretical
investigations tend to show that, in order to produce powerful
self-collimated jets, one has to rely on an accretion disc and a large
scale magnetic field. Of course, the possibility that jets arise from the
interaction between the central object magnetosphere and the disc has still
to be worked out (Camenzind \cite{came90}, Shu et al.
\cite{shu94}). Nevertheless, the study of such an interaction requires 
first the deep understanding of the interplay between accretion and ejection
processes. Hereafter, we call Magnetized Accretion-Ejection Structures
(MAES), objects where these two processes are interdependent.

There have been a number of studies of magnetized jets in the past
(BP82, Camenzind \cite{came86}, Lovelace et al. \cite{love87}, Heyvaerts \&
Norman \cite{heyv:norm}, Chiueh et al. \cite{chiu91}, Pelletier \& Pudritz
\cite{pell:pudr}, Li et al. \cite{li92}, Appl \& Camenzind
\cite{appl:came93a}, Rosso \& Pelletier \cite{ross:pell}, Contopoulos
\cite{cont95}, to cite only a few), but all these works were focused on jet
dynamics. Thus, it was not yet proved that the underlying disc could
indeed provide the required boundary conditions. 

Despite serious advances in the theory of magnetized accretion discs
driving jets (K\"onigl \cite{koni89}, Ferreira \& Pelletier \cite{FP93a},
\cite{FP93b}, \cite{FP95}, Wardle \& K\"onigl \cite{ward:koni}, Li
\cite{li}), this question has not yet been fully addressed. Indeed, in both
Wardle \& K\"onigl and Li approaches, the disc solutions were obtained by
not properly treating the disc vertical equilibrium and were directly
matched to BP82's jet solutions (see Sect. 4.2.1). Moreover, by doing so,
they were not able to specify the physical process leading plasma to change
its radial motion (accretion) into a vertical one (ejection). Ferreira \&
Pelletier (\cite{FP95}, hereafter FP95) constructed disc solutions by
taking into account all the dynamical terms, thus being able to answer this
question, as well as derive the physical conditions required to
magnetically launch jets. However, it remained to be proved that their
solutions could indeed produce super-Alfv\'enic jets.

The goal of this paper is therefore to construct global, non-relativistic
solutions for magnetically-driven jets from Keplerian discs. Our treatment
allows a smooth transition between the resistive disc and the ideal MHD
jet. This paper is organized as follows. We start by briefly recalling
the MHD equations of MAES and their main features. In particular, we
expose some general results on the physical processes that govern the
disc. Section 3 is devoted to the properties of magnetically-driven
jets and the constraints arising from their interplay with the underlying
disc. Self-similar, non-relativistic solutions that smoothly cross both
slow-magnetosonic and Alfv\'en critical points are displayed in Section
4. In the following section, we show the asymptotic behaviour of
self-similar jets and investigate analytically the reason for their
systematic behaviour. We conclude by summarizing our results in Section 6.

\section{Keplerian accretion discs driving jets}

\subsection{Magnetohydrodynamic equations for MAES}

In order to produce bipolar jets, it is natural to rely on a bipolar
topology for the large scale magnetic field. The other alternative, a
quadrupolar topology, could in principle be used too, but such a topology
could only produce weak jets (if any, see Appendix \ref{Ap:quadru}). In both  
cases, this large scale magnetic field has two distinct possible origins:
advection of interstellar magnetic field with accreting matter or
``in-situ'' production through a disc dynamo (Pudritz \cite{pudr81}, Khanna
\& Camenzind \cite{khan:came}). Both scenarii raise unanswered
questions. For example, on the degree of diffusion of the 
infalling matter, which strongly determines the field strength in the
inner regions (Mouschovias \cite{mous91}, FP95). On the other hand,
dynamo theory is still kinematic and one cannot easily infer from these
studies what would be the final stage of the magnetic topology, if the
matter feedback on the field is taken into account (Yoshizawa \& Yokoi 
\cite{yosh:yoko}). Most probably, a realistic scenario would have to take
into account advection of external magnetic field while its amplification
by the local dynamo. 

In what follows, we restrict ourselves to the bipolar topology and assume
that this field, although it has a profound dynamical influence on the
disc, is not strong enough to significantly perturb its radial
balance. Thus, the disc is supposed to be geometrically thin (its
half-width at a distance $r$ verifies $h(r)\ll r$), in a quasi
Keplerian equilibrium. The magnetohydrodynamic equations describing
stationary, axisymmetric MAES are then the following:
\begin{eqnarray}
& & \nabla \cdot\rho \vec{u}   =  0 \label{eq:mass}\\
& & \rho \vec{u}\cdot\nabla\vec{u} =  - \rho \nabla \Phi_G\; - \;\nabla P 
\;+ \;\vec{J}\times\vec{B}\; + \; \nabla\cdot \vec{T}\\
& & \vec{J}  =  {1\over \mu_o} \nabla \times \vec{B} \\
& & \eta_m J_{\phi}  =  \vec{u}_p \times \vec{B}_p \label{eq:diff}\\
& & \nabla \cdot ( {\nu'_m \over r^2} \nabla r B_{\phi})  =   \nabla
\cdot{1\over r} (B_{\phi}\vec{u}_p - \vec{B}_p\Omega r ) \label{eq:ind}
\end{eqnarray}
\noindent where $\Phi_G= - GM/(r^2 + z^2)^{1/2}$ is the gravitational
potential of the central object (disc self-gravity is neglected) and
$\vec{T}$ is a viscous stress tensor (Shakura \& Sunyaev
\cite{shak:suny73}). Both disc velocity and magnetic field were decomposed
into poloidal and toroidal components, namely $\vec{u} = \vec{u}_p + \Omega
r\vec{e_{\phi}}$ and $\vec{B} = \vec{B}_p + B_{\phi} \vec{e_{\phi}}$
respectively. A magnetic bipolar topology imposes an odd $B_{\phi}$
with respect to $z$, as well as a poloidal field 
\be
\vec{B}_p  = \frac{1}{r}\nabla a \times \vec{e_{\phi}} \ ,
\label{eq:topo}
\ee
\noindent where $a(r,z)$ is an even function of $z$. A magnetic surface,
which is a surface of constant magnetic flux, is directly labelled by
$a(r,z)= a_o(r_o,0)$. The distribution of magnetic flux through the disc is
a free (and unknown) function. We will therefore use a prescription
consistent with the complete set of equations. 

Since the magnetic field threads the disc, steady-state accretion requires
that matter diffuses through the field. As usual in astrophysics, normal
transport coefficients are far too small to account for the expected
motions. Drift between ions and neutrals, known as ambipolar diffusion,
could play such a role in accretion discs (K\"onigl \cite{koni89}, Wardle
\& K\"onigl \cite{ward:koni}). However, jets are 
observed in a wide variety of objects, which suggests that they are
produced by a mechanism independent of the disc ionisation
degree. Moreover, it is now well known that magnetized discs are prone to
instabilities (e.g. Balbus \& Hawley \cite{balb:hawl}, Tagger et
al. \cite{tagg92}, Foglizzo \& Tagger \cite{fogl:tagg}, Curry \& Pudritz 
\cite{curr:pudr}, Spruit et al. \cite{spru95}). Thus, we assume that the
disc is turbulent and that the non-linear evolution of this turbulence
provides the required anomalous transport coefficients, such as the
magnetic diffusivity $\nu_m$, resistivity $\eta_m= \mu_o \nu_m$ and
viscosity $\nu_v$ (appearing in $\vec{T}$). All these transport
coefficients should achieve a comparable level (Pouquet et
al. \cite{pouq76}). Nevertheless, we introduced in Eq.(\ref{eq:ind}) a
``toroidal'' magnetic diffusivity $\nu'_m$ to account for anisotropy with
respect to the ``poloidal'' diffusivity $\nu_m$. Indeed, the winding up of
the field due to disc differential rotation can lead to strong
instabilities and thus, to enhanced transport coefficients with respect to
the toroidal field (see FP95). 

Finally, in order to close the system, we use a polytropic approximation
\be
P = {\cal K} \rho^{\gamma}
\label{eq:gas}
\ee
\noindent with ${\cal K}$ constant along a magnetic surface, in the
isothermal case ($\gamma = 1$). Magnetized jets driven by an accretion disc
can be initially launched either by predominant magnetic effects
(magnetically-driven) or by enthalpy alone (thermally-driven). For discs
without a hot corona (cold MAES with negligible enthalpy), FP95 showed that
jets could indeed be produced through magnetic effects alone. For these
magnetically-driven jets, the vertical profil of the temperature has not a
strong influence on plasma dynamics, therefore allowing a description with
an isothermal structure. The reason lies in the fact that most of the
accretion power goes into the jets as an MHD Poynting flux. Thus,
magnetically-driven jets will always be associated with weakly dissipative
discs, where thermal effects can be neglected (see below).  

Such a cold MAES corresponds to a ``clean'' magnetic structure at the disc
surface, with just one polarity. In fact, it is likely that the magnetic
structure would consist of open field lines coexisting with small scale
loops anchored at different radii (Galeev et al. \cite{gale79}, Heyvaerts
\& Priest \cite{heyv:prie}). Such a situation would then most probably
result in the formation of a hot corona. The treatment of such a hot MAES
is postponed to future work.

The local state of a cold MAES is mainly characterized by the set of
following parameters evaluated at the disc midplane, 
\begin{eqnarray}
\varepsilon & = & \frac{h}{r} \nonumber \\
{\cal R}_m & = & \frac{r u_r}{\nu_m}  \nonumber \\
\alpha_m & = & \frac{\nu_m}{V_A h}  \nonumber \\
\mu & = & \frac{B^2}{\mu_o P} \label{eq:disc}\\
\xi & = & \frac{\dd \ln \dot M_a}{\dd \ln r}\nonumber \\ 
\Lambda & = & \frac{(\vec{J}\times\vec{B})_{\phi}} {(\nabla\cdot
\vec{T})_{\phi}} \simeq  \frac{ 2 q \mu}{\alpha_v \varepsilon} \nonumber
\end{eqnarray}
\noindent namely, the disc aspect ratio $\varepsilon \ll 1$, the magnetic
Reynolds number ${\cal R}_m$, the turbulence level parameter
$\alpha_m$, the disc magnetization $\mu$, the ejection index $\xi > 0$
and the ratio $\Lambda$ of the magnetic to the viscous torque
(FP95). This ratio depends on the magnitude $\mu$ of the field, the
magnetic shear $q$ of order unity, defined as
\be
q \equiv - \frac{h}{B_o} \left. \frac{\partial B_{\phi}}{\partial z}
\right |_{z=0}
\ee
\noindent and being a measure of the toroidal field at the disc surface
(Ferreira \& Pelletier \cite{FP93a}, hereafter FP93a) and, of course, on
the magnitude of the viscosity through the well known $\alpha_v$ parameter
(Shakura \& Sunyaev \cite{shak:suny73}). Under the sole assumption that the
disc is in Keplerian balance, angular momentum conservation implies that
\be
 1 + \Lambda \simeq {\cal R}_m \left (\frac{\nu_m}{\nu_v}\right)_{z=0}
\ee
\noindent must be verified at the disc midplane. If large scale magnetic
fields are irrelevant ($\Lambda \ll 1$), one recovers that the Reynolds
number ${\cal R}_v = r u_r/\nu_v$ is of order unity, which is required
in standard viscous discs. On the contrary, when they significantely brake
the disc, one obtains that the degree of curvature of the field lines at the
disc surface (measured by ${\cal R}_m$) depends mostly on how strong is
$\Lambda$. BP82 showed that in order to magnetically launch jets without a
hot corona, the magnetic structure needs to be bent by more than $30^o$
with respect to the vertical axis. This implies a magnetic Reynolds number
of order ${\cal R}_m \ga \varepsilon^{-1}$, therefore $\Lambda \ga
r/h$. Cold jets require then a dominant magnetic torque, imposing a
corresponding value on both the field strength and shear.

Energy conservation equation (FP93a),
\be
P_{lib} = 2 P_{rad} \; + \; 2 P_{MHD}\; + \; 2P_{th}
\ee
\noindent shows that the total available mechanical power $P_{lib}$ is
shared by the disc luminosity $P_{rad}$, the outward MHD Poynting flux
$P_{MHD}$ and the flux of thermal energy $P_{th}$ from each surface of the
disc. The liberated power is such that $P_{lib} \la P_{acc} \equiv GM\dot
M_{ae}/2 r_i$ (see FP95, Ferreira \cite{F96}), where $r_i$ is the disc
inner radius and $\dot M_{ae}$ is the accretion rate at the disc outer edge
$r_e$. Around a compact object, this accretion power is  
\be
P_{acc} \simeq 4.7\ 10^{45}\ \left( \frac{\dot M_{ae}}{\dot M_*}\right)
\left(\frac{r_i}{r_*}\right )^{-1} \ \mbox{erg s}^{-1} 
\ee
\noindent for $\dot M_* = 1 M_{\sun} yr^{-1}$, $r_* = 3r_g$, where
$r_g=2GM/c^2$ is the Schwarzschild radius. Around a protostar, this power is
\be
P_{acc} \simeq 10^{32}\ \left( \frac{\dot M_{ae}}{\dot M_*}\right)
\left (\frac{M}{.5 M_{\sun}} \right) \left(\frac{r_i}{r_*}\right )^{-1}
\mbox{erg s}^{-1} 
\ee
\noindent for $\dot M_* = 10^{-7} M_{\sun} yr^{-1}$, $r_* = 30 r_{\sun}$
(10 stellar radii for a typical T-Tauri star). Since the available energy
is stored into rotation, energy conservation can be rewritten as   
\begin{eqnarray}
2P_{rad} & \simeq & \frac{1 - \Theta}{1 + \Lambda} \ P_{lib} \nonumber \\
2P_{jet} & = & \frac{\Lambda + \Theta}{1 + \Lambda} \ P_{lib} \ ,
\label{eq:Pj}
\end{eqnarray}
\noindent where $\Theta < 1$ is the fraction of the internal energy that
was transferred in the jet as enthalpy. The total jet power, which can be
either radiated away or injected at the terminal shock (e.g. extragalactic
radio lobes), arises from the sum of the MHD Poynting flux and the plasma
thermal power. Thus, magnetized accretion discs without a hot corona
($\Lambda \simeq r/h$, $\Theta = 0$) radiate only a fraction of order $h/r$
of the jet power.

\begin{figure*}
%\picplace{10.3cm}
\psfig{figure=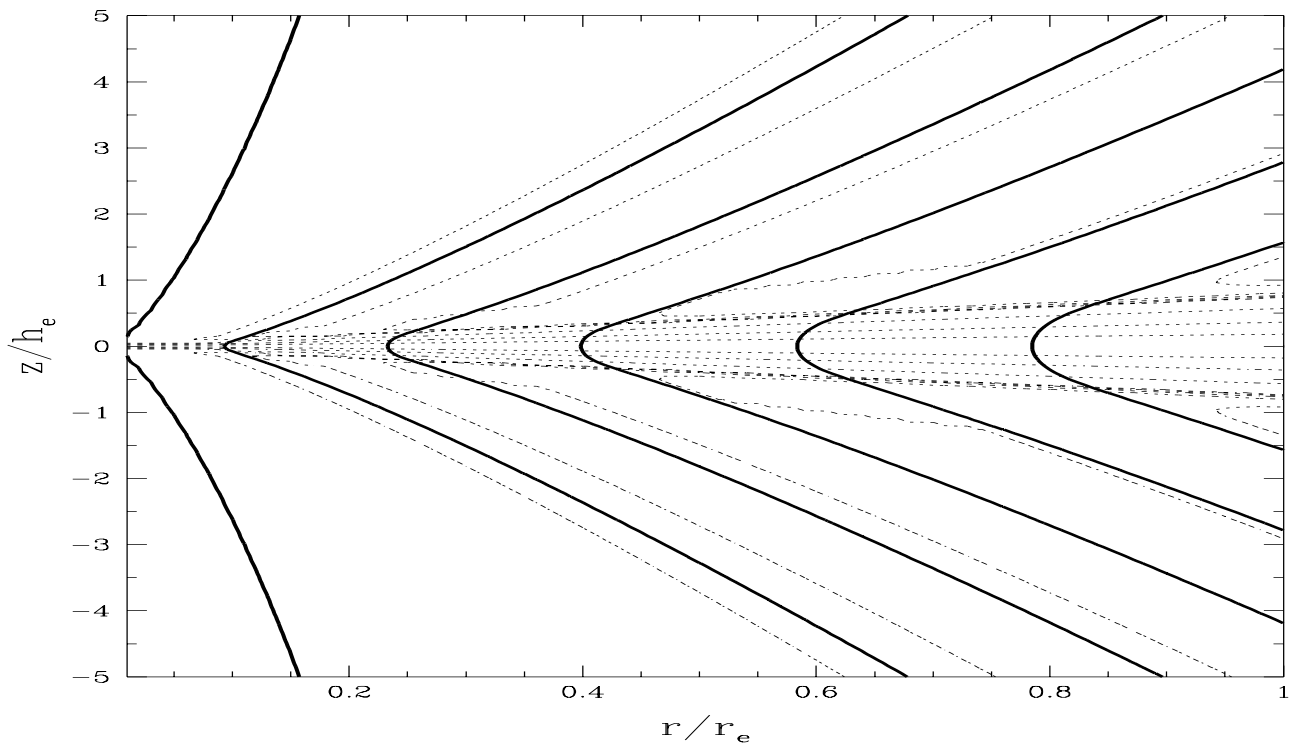,height=10.3truecm,width=18truecm}
\caption[ ]{Side view of a magnetized accretion disc driving jets, with 
$\varepsilon= 10^{-1}$, $\alpha_m=1$ and an ejection efficiency
$\xi=.004$. Plasma (dotted lines) enters the structure at its outer edge 
and is accreted with a slight converging motion through the magnetic field
(solid lines). Since the magnetic pinching force decreases vertically,
plasma reaches a layer where the pressure gradient slowly expells
it. Diffusion is still necessary at the disc surface in order to 
allow a transition between the accretion disc and the ideal MHD jet. This
transition region is approximatively one disc scale height thick. Farther 
out, plasma is frozen in a particular magnetic field line and is accelerated
through the Lorentz force. Note the scaling factor $\varepsilon^{-1}$
applied to the vertical axis.}
\label{fig:imadisc}
\end{figure*}

\subsection{From accretion to ejection}

MAES are intricate structures where accretion and ejection are
interdependent. Therefore, we will frequently make references to disc
physics and quantities in our investigation of jet physics. Let us
then, for the sake of completeness, summarize here results concerning
magnetized disc physics.

Two simultaneous processes are responsible for accretion (FP93a): (1) a
turbulent magnetic diffusivity $\nu_m$ allowing matter to steadily diffuse
through the field; (2) a dominant magnetic torque $F_{\phi}= J_zB_r -
J_rB_z$, which brakes the disc and stores into the field both angular
momentum and mechanical energy of the plasma. A local
quasi-magnetohydrostatic vertical equilibrium is achieved, the plasma
pressure gradient hardly competing with both tidal compression and vertical
magnetic pinching force. A radial magnetic tension slightly counteracts
gravity, thus leading to a sub-Keplerian rotation rate.

As a result, plasma inside the disc (see Fig. \ref{fig:imadisc}) is being
accreted, slightly converging towards the disc midplane (both $u_r$ and
$u_z$ are negative). How then is ejection achieved ? 

Ejection comes out naturally if the radial current density $J_r$ decreases
vertically on a disc scale height (Ferreira \& Pelletier
\cite{FP93b}). Indeed, in a Keplerian disc, this is the only possibility to
change the sign of the magnetic torque $F_{\phi}$. This torque must become
positive at the disc surface in order to provide a magnetic
acceleration. Such a situation requires that the counter current due to the
disc differential rotation balances the current induced by the unipolar
induction effect (FP95). If this is not fulfilled, the complete structure
is unsteady: in fact, the requirement of stationarity provides the level of
the ``toroidal'' diffusivity $\nu'_m$.   

The vertical decrease of $J_r$ leads to the following important effects:  

(1) the vertical Lorentz force decreases, allowing the plasma pressure
gradient to gently lift up matter from the disc surface;

(2) matter is azimuthally accelerated by the magnetic field, leading then to
a radial centrifugal acceleration.

\noindent Note that, although it is plasma pressure that provides a positive
vertical velocity, this process is purely magnetic. It arises naturally if
the magnetic pinching force (that must be comparable to the plasma
pressure gradient) decreases vertically. 

In order to get an insight on the jet physical processes, it is
worthwhile to project the Lorentz force parallel and perpendicular to any
poloidal magnetic surface, 
\begin{eqnarray}
F_{\phi} & = & \frac{B_p}{2\pi r} \nabla_{\parallel} I \nonumber \\
F_{\parallel} & = & - \frac{B_{\phi}}{2\pi r}\nabla_{\parallel} I
\label{eq:force}\\
F_{\perp} & = & B_pJ_{\phi} - \frac{B_{\phi}}{2\pi r} \nabla_{\perp} I \ .
\nonumber
\end{eqnarray}
\noindent Here, $I = 2 \pi r B_{\phi}/\mu_o <0 $ is the total current flowing
within this magnetic surface, $\nabla_{\parallel}\equiv (\vec{B}_p
\cdot\nabla)/B_p$ and $\nabla_{\perp}\equiv (\nabla a \cdot\nabla)/|\nabla
a|$. This shows directly that plasma is accelerated by the current leakage
through this surface ($\nabla_{\parallel} I >0$). This effect gives rise to
both a poloidal ($F_{\parallel}$) and a toroidal ($F_{\phi}$) magnetic
force, the latter providing the centrifugal force. Thus, cold jets are
better referred to as being magnetically-driven rather than
centrifugally-driven (this was also pointed out by Contopoulos \& Lovelace
\cite{cont:love}). When the current $I$ vanishes, or when it flows parallel
to the magnetic surface, no magnetic acceleration arises anymore and the
plasma reaches an asymptotic state. Note also that the way this current is
distributed across the magnetic surfaces is of great importance for the jet
transverse equilibrium ($F_{\perp}$). This shows that one has to be careful
when dealing with a particular current distribution, since it is of so
great importance in both jet collimation and acceleration (Appl \&
Camenzind \cite{appl:came93a}, \cite{appl:came93b}). 

The global current topology is linked to the ejection efficiency, which is
measured by the ejection index $\xi$ ($\xi=0$ in a viscous disc without
jet). For small ejection efficiencies ($\xi < 1/2$), ejection takes place
against a pinching vertical Lorentz force, with a positive radial
component. For those tenuous ejections, the current enters the disc at its
inner edge, flows up inside the jet and closes back along the axis. High
ejection efficiencies ($\xi > 1/2$) are achieved when matter is lifted
by both plasma and magnetic pressure gradients (positive vertical Lorentz
force), the corresponding radial Lorentz force being negative. In this
case, the current flows down the jet, enters the disc at its surface and
closes in an outer cocoon (Ferreira \cite{these}).

The set of MAES parameters (\ref{eq:disc}) is diminished by the requirement
that the overall structure is in steady-state. This implies that the flow
has to smoothly cross the usual MHD critical points it encounters (see
FP95). The constraint due to the first critical point, the
slow-magnetosonic (SM) point, allows these two current topologies to be
realized. In what follows, we will show that only tenuous ejections can
produce trans-Alfv\'enic jets.

\section{Non-relativistic, magnetically-driven jets from Keplerian discs}

\subsection{Governing equations}

Above the disc, both the decay of the turbulent diffusivity and the action
of the poloidal Lorentz force leads matter into ideal MHD state (Ferreira
\cite{F96}). In this state, matter is frozen in the field, the poloidal
motion occurring along a magnetic surface. Together with axisymmetry, this
allows to describe jets as a bunch of magnetic surfaces, nested on the
accretion disc at an anchoring radius $r_o$. In this section, we write the
well-known governing equations for self-confined jets in a form similar to
the one Pelletier \& Pudritz (\cite{pell:pudr}, hereafter PP92)
used. Written in such a form, these equations will allow us to derive
general properties and conditions for non-relativistic, magnetically-driven
jets when one takes into account the underlying disc. In Sect. 4, we will
come back to the set of equations as described in Sect. 2 and solve them
from the disc equatorial plane to the ``jet end''.

In the ideal MHD jet, Eq.(\ref{eq:diff}) becomes
\be
\vec{u}_p = \frac{\eta(a)}{\mu_o \rho} \vec{B}_p \ ,
\label{eq:up}
\ee
\noindent where $\eta(a)$ is a constant along a particular magnetic
surface. Its value, $\eta = \sqrt{\mu_o\rho_A}$, is directly evaluated at
the Alfv\'en point (labelled with the cylindrical coordinates $r_A$ and
$z_A$), where the jet poloidal velocity reaches the local Alfv\'en
speed. Equation (\ref{eq:ind}) becomes  
\be
\Omega_*(a) = \Omega - \eta \frac{B_{\phi}}{\mu_o\rho r} \ ,
\ee
\noindent and defines the rotation rate $\Omega_*$ of a magnetic
surface. Since the field lines are anchored on the accretion disc, they
rotate with roughly the same rotation rate as matter, namely $\Omega_*
\simeq \Omega_o$. Thus, one can look at this equation as providing the
amount of toroidal field required to maintain matter, which is frozen in
the field, rotating in the jet at a different rate than the field.

Using these two equations, the steady-state angular momentum conservation
equation can then be written as 
\be
\Omega_*r_A^2 = \Omega r^2 - \frac{rB_{\phi}}{\eta}
\label{eq:ang}
\ee  
\noindent where $\Omega_* r_A^2$ is the total specific angular momentum
carried away by both matter and field in a particular magnetic
surface. The radial and vertical momentum conservation equations are 
usually replaced by their projection along (Bernoulli equation) and
perpendicular (Grad-Shafranov equation, also known as transfield 
equation, Tsinganos \cite{tsin81}) to such a magnetic surface. For
polytropic flows, Bernoulli equation reads 
\be
\frac{u^2}{2} + H + \phi_G - \Omega_*\frac{rB_{\phi}}{\eta}= E(a)
\label{eq:Ber}
\ee
\noindent where $E(a)$ is the constant specific energy carried by the jet
and the jet enthalpy $H$ is defined as $\nabla H= \nabla P/\rho$. For
magnetically-driven jets (i.e. cold), this enthalpy plays no role neither
in jet formation nor in its collimation. Indeed, it is only a fraction of
order $\varepsilon^2$ of the Keplerian speed and so it will be simply dropped
in the following section (but note that the solutions in Sect. 4 are
obtained with the full set of self-similar MHD equations). As shown in
FP95, ejection of matter from the disc arises naturally due to the vertical
decrease, on a disc scale height, of the radial current density. This is a
pure magnetic process, without any help from thermal effects. On the
contrary, thermally-driven jets require a high enthalpy with a polytropic
index $\gamma$ variable along a magnetic surface, namely close to unity near
the disc to account for coronal heating and closer to the adiabatic value
further away (Weber \& Davis \cite{webe:davi}, Tsinganos \& Trussoni
\cite{tsin:trus}, Sauty \& Tsinganos \cite{saut:tsin}). This last
possibility is postponed to future work. Bernoulli equation (\ref{eq:Ber})
describes how the total energy carried by the flow is transformed into
kinetic energy, for a given magnetic configuration. Hence, one can interprete
this equation as providing the velocity matter reaches in a given
``magnetic funnel''. The shape of this funnel, or more precisely the jet
transverse equilibrium, is provided by Grad-Shafranov equation

\begin{eqnarray}
\nabla \cdot (m^2 -1) \frac{\nabla a}{\mu_o r^2} & = & \rho \left \{
\frac{\dd E}{\dd a} - \Omega \frac{\dd\Omega_*r_A^2}{\dd a}
\right. \nonumber \\   
& & +\  \left . (\Omega r^2 - \Omega_* r^2_A)\frac{\dd\Omega_*}{\dd a} 
\right \} \nonumber \\ 
& & +\  \frac{B^2_{\phi} + m^2B^2_p}{2\mu_o} \frac{\dd\ln\rho_A}{\dd a}  
\label{eq:GS}
\end{eqnarray}
\noindent where we introduced the Alfv\'enic Mach number $m^2 \equiv
u^2_p/V^2_{Ap}$.

The set of equations (\ref{eq:up}) to (\ref{eq:GS}), together with
Eq.(\ref{eq:gas}), completely describe magnetic jets where plasma pressure
plays no dynamical role. Hereafter, we describe the leading parameters of
these equations.

\subsection{An unique parameter for cold jets}

We choose to characterize our jet solutions by using parameters as defined
in the pioneering work of BP82. Since each parameter is strictly defined at
the footpoint $r_o$ of a magnetic surface, it characterizes the jet state
only locally. Thus, in a realistic 2-D case, one would have to prescribe
them for a range of anchoring radii in the disc. However, if we assume that
jets are produced from a large radial extension in the disc, then we can
look for a jet solution defined with parameters that are constants (as in
self-similar solutions), or slowly varying with the radius. Through all
this paper, we will label with a subscript ``o'' any quantity evaluated at
the disc midplane and a subscript ``A'' at the Alfv\'en surface. In
particular, $\rho_o= \rho(r_o,0)$ will be the midplane density and $u_o=
-u_r(r_o,0)$ the accretion velocity at a radius $r_o$.

The first parameter describes the magnetic configuration, namely 
\be 
\beta \equiv \frac{\dd \ln a}{\dd \ln r_o}
\ee
\noindent with $0 < \beta < 2$ (FP93a). Jets with constant $\beta$
have a magnetic field varying with a power law of the radius. BP82's
self-similar solutions were obtained with the prescription $\beta= 3/4$.  
The second parameter, defined as  
\be 
\lambda \equiv \frac{\Omega_* r_A^2}{\Omega_o r_o^2} \simeq \frac{r^2_A}
{r^2_o} 
\ee
\noindent is a measure of the magnetic lever arm that brakes the underlying
accretion disc. Since the available energy is stored in the disc as
rotational energy, a constraint on this lever arm arises through the
requirement that jets are powered with a positive energy. Indeed, the
total specific energy carried away by the (cold) magnetic structure writes
\be
E(a) = {\cal E}(a) + \Omega_*^2 r_A^2 
\ee
\noindent where ${\cal E}(a) \simeq - 3\Omega_o^2 r_o^2/2$. Jets become
free from the potential well if the ``rotator'' energy, namely $\Omega_*^2
r_A^2$, overcomes the generalized pressure ${\cal E}$ (PP92). This is
fulfilled provided $\lambda > 3/2$. The last parameter, 
\be
\kappa \equiv \eta \frac{\Omega_o r_o}{B_o}
\ee
\noindent measures the mass load on a particular magnetic surface. Since
cold jets carry away the whole disc angular momentum, one expects to find a
systematic relation between mass load and lever arm. 

These parameters must be linked to the ejection index, since it is a local
measure of the ejection efficiency. A way to find such a link is to look at
the ratio of the MHD Poynting flux to the kinetic energy flux along a
magnetic surface,   
\be
\sigma = {{- 2 \Omega_* r B_{\phi} B_p} \over {\mu_o \rho u^2 u_p}} \ .
\ee
\noindent Such a ratio measures the amount of energy stored as magnetic
energy with respect to the kinetic energy, being therefore a measure
along the jet of the acceleration efficiency. At the jet basis, identified
here as the SM-surface, the disc provides
\be
\sigma_{SM} = \frac{\dot M_a}{2\pi \rho u_z r_o^2} \frac{- B_{\phi} B_z}
{\mu_o \rho_o u_o \Omega_o h} \simeq \xi^{-1}
\ee
\noindent where we used mass ($\dd \dot M_a/\dd r = 2 \dd \dot M_j/\dd r =
4\pi\rho u_z r$) and angular momentum conservation. Thus, the ejection
index is also quite accurately the injection of energy into the jets. While
the magnetic lever arm is directly $\lambda= 1 + \sigma_{SM}/2$, the mass
load writes 
\be 
\kappa = 2\sigma_{SM}^{-1} \left| \frac{B_{\phi,SM}}{B_o}\right| \ , 
\ee
\noindent which provides $\kappa \simeq q \xi \sim \xi$ (see Appendix
\ref{Ap:Bphi}). Thus, the leading parameters for magnetically-driven jets
can be simply expressed as 
\begin{eqnarray}
\beta & = & \frac{3}{4} + \frac{\xi}{2} \nonumber \\
\label{eq:param}
\lambda & \simeq & 1 + \frac{1}{2\xi} \\
\kappa & \simeq & \xi \nonumber
\end{eqnarray}
\noindent The expression for the magnetic configuration parameter $\beta$
is general for any Keplerian disc, allowing to take into account
simultaneously magnetic and gravitational terms (FP93a, FP95). All
three parameters for cold jets can be quite accurately replaced by $\xi$:
magnetic configuration and both mass load and lever arm are therefore
tightly related. As a consequence, the parameter space for cold jets is
mostly controlled by the range of allowed ejection indices. 

By using (\ref{eq:param}), we are now able to see that any jet model that
would allow ejection with a high efficiency (namely, $\xi \ge 1$) would not
obtain free jets ($E(a) \le 0$). Nevertheless, jet solutions that 
successfully cross the SM-point with $\xi >1 $ were found to be possible
(Ferreira \cite{these}). Such a situation is therefore a transient feature,
matter failing to reach the Alfv\'en  surface and falling down to the disc
after having been ejected out.

\subsection{Constraints on the ejection index}

\subsubsection{Minimum ejection index: disc vertical equilibrium}

A look at the parameter space in FP95, derived with the sole constraint due
to the slow-magnetosonic point, indicates that there is a minimum ejection
index $\xi_{min}$. Here, we show that this minimum ejection efficiency is
related to the existence of a quasi-magnetohydrostatic (MHS) equilibrium
inside the disc. Such an equilibrium, described by
\be
\rho u_z \frac{\partial u_z}{\partial z} \simeq - \frac{\partial
P}{\partial z} - \rho \Omega^2_K z - \frac{\partial}{\partial z}
\frac{B^2_r + B^2_{\phi}}{2\mu_o }   \ ,
\ee
\noindent is obtained when plasma pressure gradient balances both
tidal compression and magnetic squeezing. This magnetic squeezing depends
on the value of the magnetic field, measured by the parameter $\mu$ (see
Eq.(\ref{eq:disc})), and on both curvature (for $B_r$) and shear (for
$B_{\phi}$) effects. When the ejection efficiency decreases, the lever arm
increases and so does the magnetic compression due to field curvature. The
only way to find out another equilibrium state is then to decrease the
value of the field itself (that is, $\mu$). However, this results in an
enhancement of the magnetic squeezing due to shear.   

This can be qualitatively shown like follows (for more details, see
Appendices \ref{Ap:Bphi} and \ref{Ap:min}). The induction equation
(\ref{eq:ind}) can be written in the disc as 
\be
\eta'_m J_r \simeq \eta'_oJ_o \;+\;r\int_0^z\!\! \dd z\vec{B}_p\cdot
\nabla\Omega \;-\; B_{\phi}u_z 
\label{eq:Bphi}
\ee
\noindent where the last term describes the effect of advection. In the
resistive disc, this term is negligible in comparison with the others. Thus,
the toroidal field at the disc surface, 
\be
B^+_{\phi} = B^{J_o}_{\phi} + B^{\nabla \Omega}_{\phi}
\ee
\noindent is mainly the sum of two contributions, $B^{J_o}_{\phi} = -
\mu_oJ_o h$ due to the unipolar induction effect and
$B^{\nabla\Omega}_{\phi}$, the counter current (of positive sign) provided
by the disc differential rotation. If the disc were rigidly rotating (i.e.,
a Barlow Wheel), it would generate the first contribution but not the
second one, therefore producing no jet. When $\mu$ decreases, the magnitude
of this counter current decreases, thereby increasing the toroidal field at
the disc surface and with it, the pinching effect of the corresponding
magnetic pressure gradient.  

There is thus a $\xi_{min}$ below which no quasi-MHS equilibrium can be
found anymore. The exact value of this minimum ejection index is 
impossible to find analytically, since it depends on a subtle
equilibrium between terms that are all of the same order of magnitude. When
such equilibrium is crudely treated, one can in principle still obtain a
matching with jet solutions but the disc would certainly not survive an
overwhelming magnetic squeezing.

In Section 4, we show the parameter space obtained for self-similar
solutions, thus providing a numerical value for $\xi_{min}$. Because
self-similarity does not influence solutions close to the Keplerian disc
but allows instead to take into account all dynamical terms, we believe
that this value is general. Since what limits the minimum ejection index is
the increasing toroidal field, one can only significantly lower
$\xi_{min}$ by decreasing the value of $B^{J_o}_{\phi}$, that is, by
lowering $J_o$. Therefore, only discs where both viscosity and magnetic
torques are relevant ($\Lambda \la 1$, requiring a hot corona to produce
jets), should allow ejection indices much smaller than those displayed
here.

\subsubsection{Maximum ejection index: jet acceleration}

Conservation of angular momentum (Eq.(\ref{eq:ang})) implies that once a
large scale magnetic field steadily brakes the disc with a significant
torque, matter must be accelerated up to the Alfv\'en point. This severily
constrains the mass load in the jet, thus providing an upper limit on the
ejection index. When gravity becomes negligible, Bernoulli equation can be
rewritten (FP93a) as 
\be
\alpha^2 \equiv \frac{u_p^2}{\Omega_o^2 r_o^2} = (1-g^2)\frac{r^2}{r_o^2}
- 3 
\ee
\noindent where the function $g$, defined as $\Omega = \Omega_* (1- g)$,
can be expressed as
\be
g = \frac{m^2}{m^2 -1}\left ( 1 - \frac{r^2_A}{r^2} \right ) \ .
\ee
\noindent It measures the discrepancy between the angular velocities of
matter and magnetic surface on which matter flows (PP92). At the jet basis,
its value is approximately zero, then it increases and reaches unity for
highly super-Alfv\'enic jets. The Alfv\'enic Mach number becomes then
simply  
\be
m^2 = \alpha \kappa \frac{B_o}{B_p} \ .
\label{eq:m2}
\ee
\noindent A necessary condition for trans-Alfv\'enic jets is obtained by
requiring that the above expression reaches or becomes much bigger than
unity. At this stage, it is convenient to introduce the ratio, measured at
the Alfv\'en point, of the rotation velocity of the magnetic surface to the
poloidal Alfv\'en speed (Michel \cite{mich}, PP92)
\be
\omega_A \equiv \frac{\Omega_* r_A}{V_{Ap,A}} = \kappa \lambda^{1/2}
\frac{B_o}{B_{p,A}} \ . 
\label{eq:omA}
\ee
\noindent This fastness parameter is a useful quantity since it encloses
an information that we cannot have without solving first the whole
structure, namely the angle of the poloidal magnetic field with respect to
the vertical axis. This ratio, which must verify 
\be
\omega_A > \left| \frac{B_{\phi}}{B_p} \right|_A \ ,
\ee
\noindent will allow us to derive a general condition for trans-Alfv\'enic
jets, in two extreme cases.  

For ``powerful'' jets (high $\sigma_{SM}$), matter is expelled off the disc
carrying almost no angular momentum. Acceleration takes place only if
$\sigma_A < \sigma_{SM}$, with $\sigma_A \simeq 2 \omega_A^2$. Thus,
these jets require 
\be
\omega_A^2 < \frac{1}{2\xi} \ .
\label{eq:c1}
\ee
\noindent This corresponds to the extreme case where it takes almost no
power from the magnetic structure to accelerate plasma up to the Alfv\'en
surface. Such a configuration could be obtained in the low mass load limit.

For ``weak'' jets (low $\sigma_{SM}$), the Alfv\'enic Mach number has to reach
at least unity. At the Alfv\'en point, Bernoulli equation writes
$\alpha_A^2= (1 - g_A^2)\lambda -3$. This shows that matter reaches its
maximum velocity when $g_A = 0$, namely when the toroidal field is almost
zero. Using Eq.(\ref{eq:m2}), it can be seen that these jets require 
\be
\omega_A^2 = \frac{\lambda}{\alpha_A^2}  > \frac{\lambda}{\lambda -
3} \ . 
\label{eq:c2}
\ee
\noindent This general condition shows that magnetically-driven jets are
fast rotators ($\omega_A > 1$) in order to successfully reach the
Alfv\'en surface (BP82, PP92, Rosso \& Pelletier \cite{ross:pell}). This is
required even in this case, where matter barely reaches it despite the
almost complete transfer of energy from the magnetic structure. High mass
loads could give rise to such jets.

Compiling criteria (\ref{eq:c1}) and (\ref{eq:c2}), one gets the necessary
condition for trans-Alfv\'enic jets
\be
\frac{1 + 2\xi}{1- 4\xi} < \omega_A^2 < \frac{1}{2\xi} \ ,
\ee
\noindent which must be satisfied at each magnetic surface.  Here, we get a
much more precise constraint since it requires 
\be
\xi < \frac{\sqrt{13}-3}{4} \simeq 0.15 \ .
\ee
\noindent This implies that higher ejection efficiencies are completely
inconsistent with a steady trans-Alfv\'enic regime. Note that this value
should be viewed as an upper limit for $\xi$, the maximum ejection index
$\xi_{max}$ being in fact smaller. This is due to the weak constraint we
used in the high-$\sigma_{SM}$ case. Contopoulos (\cite{cont95}) proposed a
simple jet model where the magnetic field has only a toroidal component. In
such a case (obtained in the limit $\kappa$ very large), the Alfv\'en
singularity disappears and the driving mechanism for jet acceleration is
the magnetic pressure gradient. We would like to stress that such a
configuration is, in principle, possible as a transient outburst from thin
discs. Indeed, FP95 showed that magnetically-driven jets can be viewed as
being either ``centrifugally-driven'' for small ejection indices ($\xi <
1/2$, the vertical component of the Lorentz force pinching the disc) or
``magnetic pressure-driven'' for high ejection indices ($\xi > 1/2$, the
toroidal magnetic pressure lifting matter up, consistent with the large
$\kappa$ limit used). However, only the first situation allows a steady
state with respect to the Alfv\'en surface. The possibility remains that
such a configuration would be steadily produced by a thick disc. In the
next section, we compute global self-similar solutions, derive both
$\xi_{min}$ and $\xi_{max}$ numerically and obtain the full parameter space
for cold MAES.

\section{Trans-Alfv\'enic, global disc-jet solutions}

\subsection{The self-similar ansatz}

In order to construct global solutions from the disc equatorial plane up
to a jet asymptotic regime, the full set of non-linear, partial
differential equations (\ref{eq:mass}) to (\ref{eq:ind}) along with
Eq.(\ref{eq:gas}) must be solved. Since MAES are intrinsically 2-D, this
requires either a numerical approach (not yet available) or some method to
reduce this set to a set of ordinary differential equations (but enforcing
some symmetry to solutions). This is what allows self-similarity, which is
the reason why it has been used by many authors (see Tsinganos et
al. (\cite{tsin96}) for a review).

We look then for self-similar solutions so that any quantity $Q$ can be
written 
\be
Q(r,z) = Q_e \left(\frac{r}{r_e}\right)^{\alpha_Q}f_Q(x) \ ,
\ee
\noindent where $r_e$ is the external radius of the disc (a standard
viscous disc is probably established from $r_e$ to larger radii). The 
self-similar variable is chosen to be $x \equiv z/h(r)$, consistent with
the gravity field at the disc neighbourhood, but also everywhere since
$h(r)= \varepsilon r$. Such a form, $h \propto r$, is expected to arise in a
Keplerian disc pervaded by a large scale magnetic field (FP93a, FP95). The
magnetic flux function is chosen to be
\be
a(r,z) = a_e \left(\frac{r}{r_e}\right )^{\beta} \psi(x)
\ee
\noindent where $\psi(-x)=\psi(x)$ and the pressure prescription such that
${\cal K}(a)= {\cal K}_e (r_o/r_e)^{\alpha_{\cal K}}$, with
$\alpha_{\cal K} = (1 - \gamma)(2\beta - 3) - 1$, the polytropic index
$\gamma$ remaining a free parameter. Since we focus on cold jets, its value
has no importance and we choose $\gamma=1$ for convenience (see discussion
after Eq.(\ref{eq:gas})).

The set of PDEs can be separated into two sets, the first consisting of
algebraic equations between the indices $\alpha_Q$ and $\beta$, the second
of non-linear ODEs on the functions $f_Q$, $\psi$ and their derivatives
(see Appendix B in FP95). Within such a prescription, all the dynamical
terms can be included, allowing henceforth a complete study of the physics
of ejection from Keplerian discs. However, only non-relativistic jets can
be studied within the same self-similarity prescription. Indeed,
relativistic speeds enforce a scaling with the speed of light (Li et
al. \cite{li92}), which is inconsistent with disc physics.

Since the boundary values $f_Q(0)$ at the disc midplane are known, we
just have to integrate the set of ODEs from $x=0$ to infinity. Once we have
a global solution as a function of $x=z/\varepsilon r$, it is
straightforward to obtain the variation of any quantity along any magnetic
surface $a=constant$ with 
\be
Q(x(a)) = Q_o f(x) \psi(x)^{-\alpha_Q/\beta}
\ee
\noindent where $Q_o = Q_e(r_o/r_e)^{\alpha_Q}$ (note the use of the notation
$x(a)$ in this case). The main difficulties come from the presence of three
critical points, whenever the plasma velocity $V\equiv
\vec{u}\cdot\vec{n}$, where 
\be
\vec{n} \equiv  \frac{\vec{e_z} - \varepsilon x\vec{e_r}}{\sqrt{1 +
\varepsilon^2 x^2}}  \ ,
\label{eq:n}
\ee
\noindent equals the three usual MHD phase speeds: slow-magnetosonic, Alfv\'en
and fast-magnetosonic, in the direction $\vec{n}$. Self-similarity not only
modifies the definitions of these phase speeds, but constrains also the
direction of propagation where critical points appear (see FP95 for more
precisions). Hence, the critical velocity $V$ is roughly $u_z$ at the disc
neighbourhood where the flow becomes super-SM, and $u_r$ much farther
away.

\begin{figure}
%\picplace{7cm}
\psfig{figure=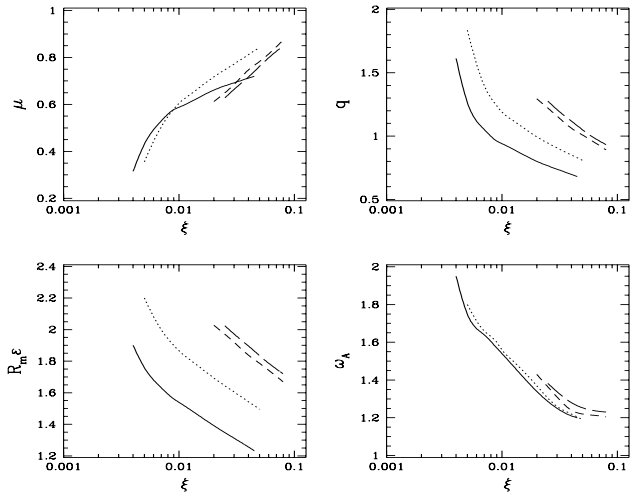,height=7truecm,width=8.8truecm}
\caption[ ]{MAES parameter space for an isothermal structure with
$\alpha_m=1$, for various disc aspect ratios: $\varepsilon = 10^{-1}$
(solid line), $10^{-2}$ (dotted line), $10^{-3}$ (short-dashed line) and
$7\ 10^{-4}$ (long-dashed line). The disc magnetization $\mu$ and the
magnetic Reynolds number ${\cal R}_m$ were obtained as regularity
conditions, in order to get respectively trans-SM and trans-Alfv\'enic
solutions. The other two parameters, the magnetic shear at the disc surface
$q$ and fastness parameter $\omega_A$ are calculated. As $\xi$ decreases,
$\mu$ must decrease in order to balance the increase in magnetic shear $q$
and curvature ${\cal R}_m \varepsilon$ (see Sect. 3.3.1). It is noteworthy
that, although the disc parameters change a little, the main effect of
decreasing $\varepsilon$ is to shift the range of allowed $\xi$ to higher
values. We confirm here that trans-Alfv\'enic jets require $\omega_A \ga 1$
(PP92, Rosso \& Pelletier \cite{ross:pell}).}      
\label{fig:para3}
\end{figure}

\subsection{MAES parameter space}

\subsubsection{The slow-magnetosonic point}

We obtain trans-SM solutions by adjusting the strength of the field for a
given set of parameters (FP95). Thus, for $\mu$ bigger than a critical
value $\mu_c$, the magnetic compression is too strong and leads to a
vanishing density (hence infinite vertical acceleration). When $\mu <
\mu_c$, the magnetic compression is insufficient and too much mass is
expelled off, which eventually falls down again. 

Thus, the smooth crossing of the first critical point is directly related
to the disc vertical balance. Following Li (\cite{li}), we can express that
a necessary (but not sufficient) condition for stationarity is 
\be
P_o \ga \left .\frac{B^2_r + B^2_{\phi}}{2\mu_o} \right|_+
\ee
\noindent where the subscript ``+'' refers to the disc surface. In terms of
MAES parameters, this condition for cold jets becomes
\be
\mu \la \frac{8 - \alpha_m^2{\cal R}_m^2\varepsilon^2 f^2_+}{4 {\cal R}_m^2
\varepsilon^2}  
\label{eq:vertmu}
\ee
\noindent where the function $f_+$ is of order unity (see Appendix
\ref{Ap:Bphi}). This implies $\mu < 3/2$, which shows that jets cannot be
produced from magnetically dominated discs. Since plasma pressure plays
such an important role in sustaining the disc, only structures close 
to equipartition are steady (FP95).

The minimum ejection index $\xi_{min}$ is found by requiring that all 
solutions become super-SM. We found $\xi_{min} = 0.004$ for an isothermal
disc with $\varepsilon=0.1$ and $\alpha_m=1$. If the disc aspect ratio
$\varepsilon$ (and also the turbulence level $\alpha_m$, see below)
decreases, the influence of advection in Eq.(\ref{eq:Bphi}) also decreases
and the radial current density $J_r$ is smaller, hence providing a smaller
value of the toroidal field at the disc surface ($f_+$, see Appendix
\ref{Ap:min}). As an example, we obtain $f_+ \simeq 0.6, 0.5, 0.45$ for
$\varepsilon= 0.1, 0.01, 0.001$ respectively. The magnetic compression
decreasing, the range of  allowed ejection efficiencies is shifted to
higher values (see Fig. \ref{fig:para3}). Moreover, by using an isothermal
prescription for the disc temperature, we have underestimated the
``lifting'' efficiency of the plasma pressure gradient. Thus, $\xi_{min}=
0.004$ should be seen as the lower limit for cold jets from Keplerian
discs. However, since $\xi < 0.15$ is a general constraint, one can expect
to find that below a certain value of $\varepsilon$ (and $\alpha_m$),
no steady-state solutions can be found anymore.

The work presented here strongly differs from the work done by Wardle \&
K\"onigl (\cite{ward:koni}) and Li (\cite{li}). These authors found
no limiting $\xi$ and were thus able to obtain jet solutions with an
enormous lever arm $\lambda$, and an arbitrarily small mass load
$\kappa$. Here, it has been found that such a situation is impossible to
achieve in steady-state. This disagreement comes from their different
treatment of the highly sensitive disc vertical equilibrium. Indeed, Wardle \&
K\"onigl investigated the vertical structure at a given radius and made the
assumption (valid for the jet) $\rho u_z =constant$ in order to deal with the
mass conservation equation. Li used a self-similar approach, but did not
make a smooth transition between the disc and its jets, crudely matching
the disc MHS solutions to jet ones. He however found that the mass load was
very sensitive to the plasma rotation rate at the disc midplane $\Omega_o$:
the smaller $\Omega_o$ the smaller $\kappa$. This can now be understood
with what was previously said. Indeed, $\Omega_o = \Omega_k(1 - \omega_o)$
with $\omega_o \simeq \mu {\cal R}_m\varepsilon^2/2 =
o(\varepsilon)$. Thus, a smaller $\Omega_o$ is achieved with a bigger
$\varepsilon$, which implies a smaller ejection efficiency (hence, $\kappa
\simeq \xi$).

\subsubsection{The Alfv\'en point}

In order to get trans-Alfv\'enic solutions, we adjust the magnetic Reynolds
number ${\cal R}_m$, which controls the bending of the poloidal field lines
at the disc surface. For ${\cal R}_m$ bigger than a critical value ${\cal
R}_c$, the bending is too strong and leads to an overwhelming centrifugal
effect, such that $\Omega r^2 > \Omega_* r_A^2$ (i.e., $B_{\phi}$ becomes
unphysically positive). On the contrary, if ${\cal R}_m < {\cal R}_c$, the
centrifugal effect cannot balance the magnetic tension, which leads to an
unphysical closure of the magnetic surfaces ($B_r$ becomes negative). At
each trial for ${\cal R}_m$ one has to find another critical $\mu_c$ that
allows a super-SM solution. Note that fixing ${\cal R}_m$ is the same as
fixing the accretion velocity at the disc midplane, for any given magnetic
diffusivity. We have the freedom to do it because this accretion velocity
is also controlled by the magnetic field. If we were in a situation where
accretion is mainly due to the viscous stress, then ${\cal R}_m$ would
also be given.

\begin{figure}
%\picplace{7cm}
\psfig{figure=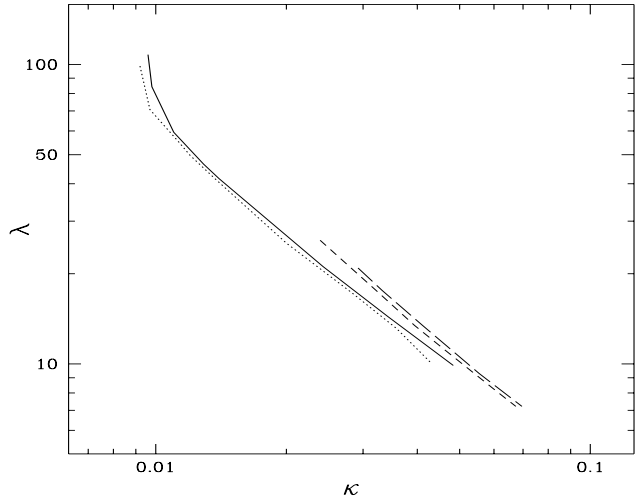,height=7truecm,width=8.8truecm}
\caption[ ]{Parameter space for cold jets launched from an isothermal 
accretion disc, with $\alpha_m =1$ and for various disc aspect ratios:
$\varepsilon = 10^{-1}$ (solid line), $10^{-2}$ (dotted line), $10^{-3}$
(short-dashed line) and $7\ 10^{-4}$ (long-dashed line). Due to the SM
constraint, we get here a smaller parameter space as in BP82. As showed,
the jet parameters do not significantly vary with any parameter but
$\xi$.}
\label{fig:para2}
\end{figure}

Figure \ref{fig:para2} shows the parameter space for cold jets in the
$\kappa$-$\lambda$ plane for $\alpha_m=1$. Solutions with $\kappa > 0.1$ do
not allow super-Alfv\'enic jets (BP82, Wardle \& K\"onigl
\cite{ward:koni}). We found here that $\xi_{max} = .08$, the maximum
Alfv\'enic Mach number reached by the flow being barely bigger than
unity. For disc aspect ratios below $5\ 10^{-4}$, no steady-state solutions
are found, $\xi_{min}$ becoming equal to $\xi_{max}$. Thus, trans-Alfv\'enic
jets are produced for discs where $5\ 10^{-4} < \varepsilon \le 10^{-1}$,
the upper bound arising from the thin-disc approximation. 

Using Eq.(\ref{eq:up}), we can express the density at the Alfv\'en point as
a function of the disc midplane density and MAES parameters, namely 
\be
\frac{\rho_A}{\rho_o} \simeq \frac{\alpha_m^2}{4} {\cal R}^2_m
\varepsilon^4 \xi^2 \ .
\label{eq:rhoA}
\ee
\noindent This relation is local and independent of any boundary condition
that could constrain the jet behaviour. Since we approximately obtain
numerically such a value for the Alfv\'en density, as well as a consistency
with all our general requirements of Sect. 3.3.2, we believe that
self-similarity doesn't affect the solutions up to this point. Therefore,
our parameter space should be viewed as general for cold jets launched from
Keplerian discs. 

\begin{figure}
%\picplace{7cm}
\psfig{figure=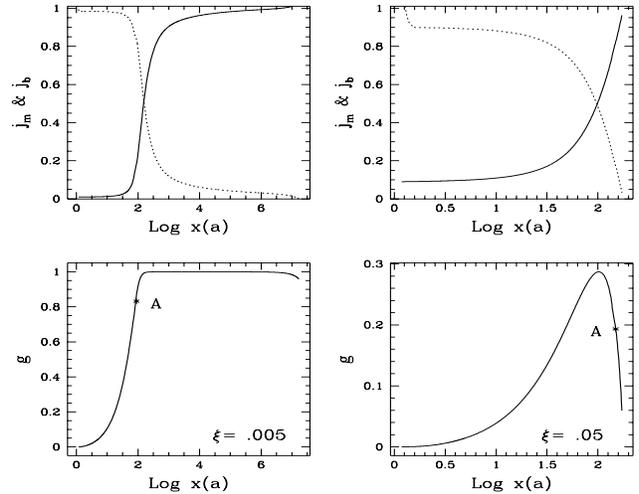,height=7truecm,width=8.8truecm}
\caption[ ]{Low-efficiency ($\xi=0.005$, left pannels) and high-effi\-ciency
($\xi=0.05$, right pannels) trans-Alfv\'enic jets for $\varepsilon=
10^{-2}$, $\alpha_m=1$. The upper pannels show the angular momentum
transfer from the field ($j_b$, dashed line), to the plasma ($j_m$, solid
line). At the disc surface ($x=1$), all the specific angular momentum $\Omega_*
r^2_A$ is carried by the field, but it is afterwards completely transferred
into the plasma. Below, the corresponding function $g= 1 -\Omega/\Omega_*$
is displayed for both cases. The cross labelled with ``A'' tells the
position of the Alfv\'en surface. In the low-efficiency limit, the jet
reaches it with $g \sim 1$ ($\omega_A = 1.8$), most of the power being
still carried by the field. In the high-efficiency case, the flow requires
already almost all the power to reach Alv\'en speeds ($\omega_A = 1.19$).} 
\label{fig:jv2}
\end{figure}

At this stage, one cannot tell with precision what will be the
asymptotic state of the jet (unless of course by propagating the solution,
as we will do next section). However, by using general arguments based
on the work of Heyvaerts \& Norman (\cite{heyv:norm}), one can give
clues on the degree of collimation achieved, according to the amount of
current $I_A$ 
\be
\frac{I_A}{I_{SM}} = g_A (1 + 2\sigma_{SM}^{-1} ) \simeq g_A
\ee
\noindent still available at the Alfv\'en point. Hereafter, we make a
distinction between two kinds of jets, depending upon their state at the
Alfv\'en surface (see Fig. \ref{fig:jv2}).

Current-carrying jets ($g_A \la 1$) reach the Alfv\'en surface while most
of the angular momentum is still stored in the magnetic structure. They are
powerful (corresponding here to ``low'' ejection indices $0.004 \le \xi \la
0.025$) and could reach highly super-Alfv\'enic speeds ($m^2 \gg 1$), with
$\Omega \simeq \Omega_* r_A^2/r^2$ and a maximum poloidal velocity
\be
u_{p,max} = \Omega_o r_o (2\lambda -3)^{1/2} \ .
\ee
\noindent Such jets could allow a cylindrical collimation if this current
does not vanish completely.

Current-free jets ($g_A \ll 1$) become super-Alfv\'enic at the expense of
almost all the available current. They correspond to ``high'' ejection
efficiencies $0.025 \la \xi \le 0.08$ (for $\alpha_m =1$, see
below). According to Heyvaerts \& Norman, only parabolic collimation could
be asymptotically achieved. These jets could reach moderate speeds ($m^2
\ge 1$), with $\Omega \la \Omega_*$ and a maximum poloidal velocity
\be
u_{p,max} = \Omega_o r_o (\lambda -3)^{1/2} \ .
\ee 

\begin{figure}
%\picplace{7cm}
\psfig{figure=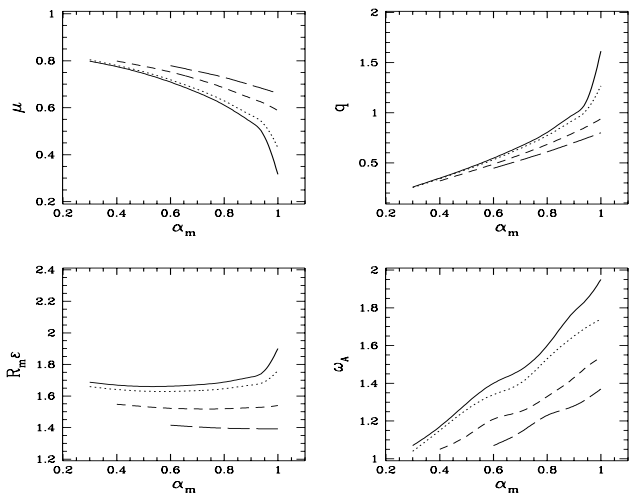,height=7truecm,width=8.8truecm}
\caption[ ]{Influence of the turbulence parameter $\alpha_m$ on other MAES
parameters (see Fig. \ref{fig:para3}), for $\varepsilon= 10^{-1}$ and
various ejection indices: $\xi= 0.004$ (solid line), 0.005 (dotted line), 0.01
(short-dashed line) and 0.02 (long-dashed line). The minimum level of MHD
turbulence is limited by the value of the induced toroidal field that 
allows trans-Alfv\'enic jets ($\omega_A >1$). The maximum level is
arbitrarily fixed to unity. Note that for fixed $\alpha_m$, the fastness
parameter $\omega_A$ grows with decreasing $\xi$ (see text).}
\label{fig:para5}
\end{figure}

\subsubsection{Influence of the magnetic diffusivity}

In the above two sections, we restricted ourselves to discs with
$\alpha_m=1$. Figure \ref{fig:para5} shows that it has a profound effect on
the fastness parameter $\omega_A$: at constant $\xi$, the smaller
$\alpha_m$ the smaller $\omega_A$. Since trans-Alfv\'enic jets require
$\omega_A > 1$, we obtain that there is a minimum turbulence level
required. Because usual dimensional arguments restrict $\alpha_m$ to
unity, we conclude that steady state MAES require an MHD turbulence with
$0.1 < \alpha_m \la 1 $. 

Equation (\ref{eq:omA}) provides, for small ejection indices (Pelletier et
al. \cite{pell96}), 
\be
\omega_A \ga \frac{q}{4}\xi^{-1/2} = \frac{\alpha_m}{8} {\cal R}_m
\varepsilon \mu^{-1/2} \xi^{-1/2} \ .
\ee 
\noindent The fastness parameter $\omega_A$ strongly depends on both the
shear parameter $q$ (henceforth on $\alpha_m$) and the ejection index
$\xi$. The highest value of $\omega_A$ (hence, of $g_A$) will be achieved with 
$\alpha_m=1$ and the smallest value of $\xi$. Since the disc vertical
equilibrium limits the latter, $\omega_A$ is therefore also limited
($1 < \omega_A < 2$, see Figs. \ref{fig:para2} and \ref{fig:para5}).

As $\alpha_m$ decreases, the influence of plasma on the field grows: the
counter current due to the disc differential rotation  increases, as well
as the effect of advection (see Appendix \ref{Ap:min}). Thus, the toroidal
magnetic field at the disc surface decreases, decreasing accordingly the
magnetic torque. This has two consequences. First, the mass load is
slightly lowered (see Eq.(\ref{eq:rhoA})), providing a higher Alfv\'en
speed at the Alfv\'en surface and a lower fastness parameter. Second, the
jet is less powerful ($\sigma_{SM}$ decreases) and the magnetic force
($F_{\perp}$, see Eq.(\ref{eq:force})) is less efficient in opening the
jet: $r_A/r_o$ decreases while $z_A/r_o$ increases. In order to accelerate
plasma up to the Alfv\'en surface, the magnetic structure has to provide
more power (through $F_{\phi}$ and $F_{\parallel}$), thereby reducing the
available current flowing inside a given magnetic surface ($g_A$
decreases). Therefore, powerful jets with current still available at the
Alfv\'en surface require a high diffusivity level ($\alpha_m \sim 1$).

We can now generalize our understanding of the asymptotic behaviour of
super-Alfv\'enic jets by using $\omega_A$ instead of $\xi$. Indeed,
although the jet parameters are mostly described by $\xi$, the jet
asymptotic behaviour is drastically sensitive to $\omega_A$.

\section{Asymptotic behaviour of self-similar jets}

\subsection{The fate of self-similar, non-relativistic jets}

\begin{figure}
%\picplace{7cm}
\psfig{figure=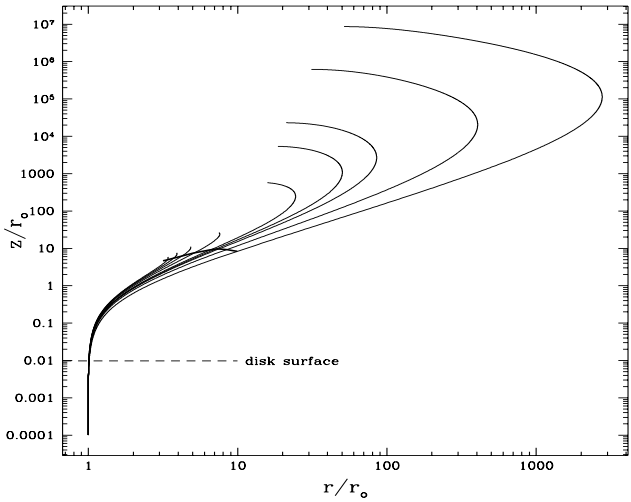,height=7truecm,width=8.8truecm}
\caption[ ]{Poloidal magnetic field lines for $\varepsilon= 10^{-2}$ and
$\alpha_m=1$, for $\xi = 0.05$, 0.04, 0.03, 0.02, 0.012, 0.01, 0.009, 0.007
and 0.005 (the maximum radius increases with decreasing ejection
index). The thick line connects the position of the Alfv\'en point for each
solution.  Note that in the range of allowed ejection indices for
$\alpha_m=1$, the appearance of jets remains quite variable: for an
anchoring radius $r_o = 10^{-1}$ AU from a young star, jets can propagate
from 1 to $10^6$ AU of the central source, with a maximum radius ranging
from 1 to approximately 300 AU. Note the logarithmic scales: small $\xi$
jets recollimate with angles smaller than one degree (Fig.\ref{fig:jv4}).}
\label{fig:psi}
\end{figure}

\begin{figure}
%\picplace{4cm}
\psfig{figure=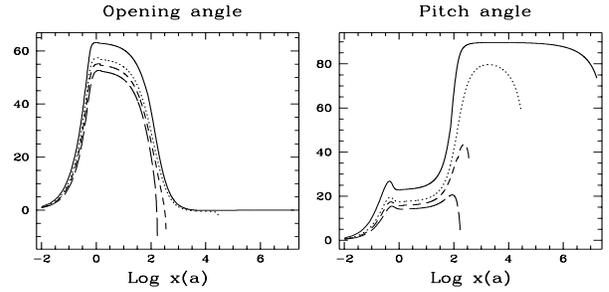,height=4truecm,width=8.8truecm}
\caption[ ]{Jet opening angle ($\theta = \arcsin (B_r/B_z)$, left pannel) and
pitch angle ($\arctan(-B_{\phi}/B_p)$, right pannel) in degrees, for 
$\varepsilon= 10^{-2}$, $\alpha_m=1$ and $\xi= 0.005$ (solid line), 0.01
(dotted line), 0.02 (short-dashed line) and 0.05 (long-dashed line). The jet
opening angle is positive when the jet widens, negative when it undergoes a
recollimation. While jets with high ejection indices refocuse with an angle
of order $-10^o$, the others are almost cylindrical, with an angle
smaller than $1^o$ (but negative).}
\label{fig:jv4}
\end{figure}

\begin{figure}
%\picplace{7cm}
\psfig{figure=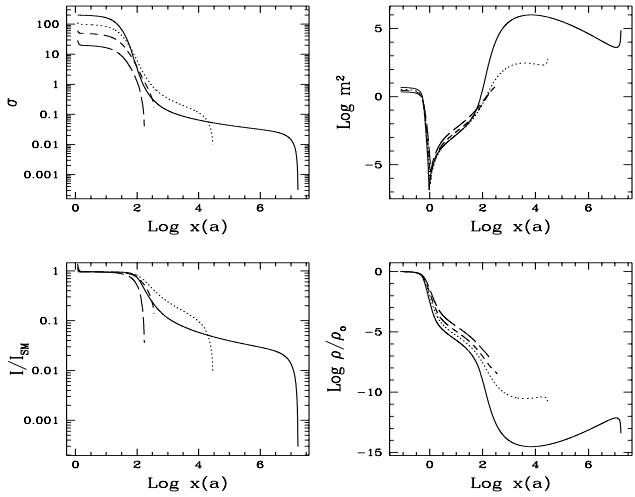,height=7truecm,width=8.8truecm}
\caption[ ]{Ratio $\sigma$ of the Poynting flux to the kinetic energy flux,
logarithm of the Alfv\'enic Mach number $m^2$, total current $I$ flowing
within the magnetic surface (normalized to the current $I_{SM}$
provided at the jet basis), and logarithm of the jet density along any
magnetic surface, for various ejection indices. These curves were
obtained for the same parameters as in Fig. \ref{fig:jv4}. The jet density
is normalized to the disc midplane density $\rho_o = \dot M_a(r_o)/8 \pi q
\mu \varepsilon^2 \Omega_o r_o^3$, at the radius $r_o$. The jet
acceleration depends on how much a magnetic surface widens ($m^2=
\rho_A/\rho$), which is possible only if the current decreases. Here, this
acceleration is very efficient, the magnetic structure feeding plasma with
almost all its power ($\sigma \ll 1$).} 
\label{fig:jv3}
\end{figure}

\begin{figure}
%\picplace{7cm}
\psfig{figure=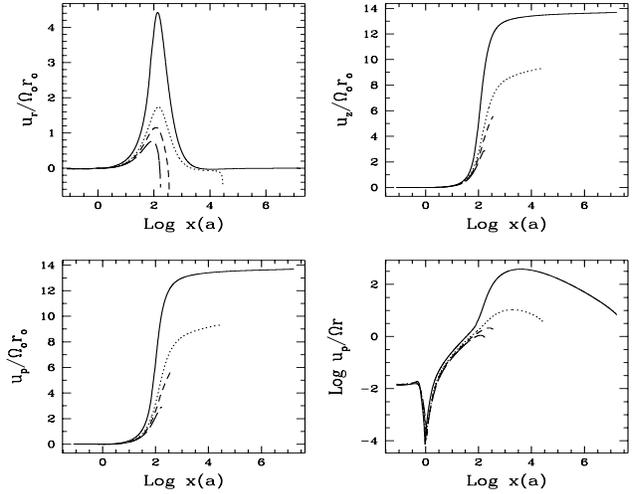,height=7truecm,width=8.8truecm}
\caption[ ]{Components of the jet poloidal velocity $\vec u_p$ and logarithm of
the ratio of the poloidal to the azimuthal velocity, measured along a
magnetic surface for various ejection indices (see Fig. \ref{fig:jv4}). For
these typical solutions, the jet always reaches its maximum velocity (see
text), mainly as a vertical component. For small ejection indices, a full
relativistic treatment should be used for jets around compact
objects. Inside the disc, matter is being accreted with a velocity of order
$\varepsilon$ the Keplerian velocity. The turning point ($u_r=0 $) occurs
roughly at the disc surface ($x=1$, see FP95).}
\label{fig:jv1}
\end{figure}

When integrated up to ``infinity'', all our solutions display the same
behaviour: after an initial widening, the magnetic surfaces reach a maximum
radius $r_t$ and then start to bend towards the jet axis (see Figs.
\ref{fig:psi} and \ref{fig:jv4}). The acceleration efficiency is very
high, since the magnetic structure converts almost all the MHD Poynting
flux into kinetic power ($\sigma \ll 1$), allowing therefore matter to
reach its maximum velocity (see Figs. \ref{fig:jv3} and \ref{fig:jv1}). 
They finally stop at a finite distance, with a cylindrical radius larger
than the Alfv\'en radius $r_A$ (Figs. \ref{fig:psi} and \ref{fig:jv4}). We
are then faced to the following obvious questions: Why do our solutions
stop ? and why do they always recollimate ?  

\begin{figure}
%\picplace{7cm}
\psfig{figure=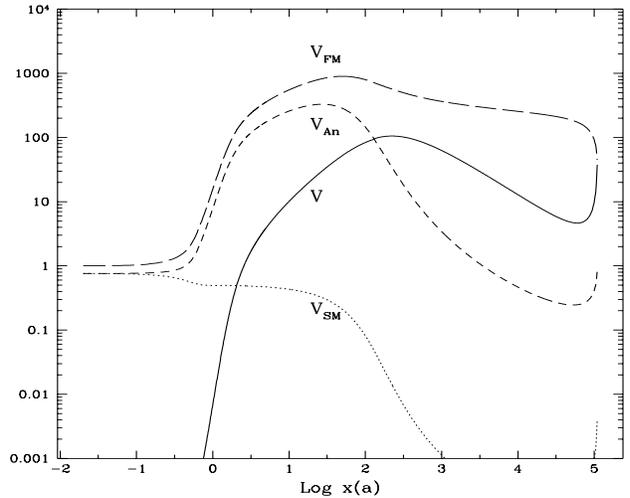,height=7truecm,width=8.8truecm}
\caption[ ]{Characteristic velocities in units of the disc sound speed
$\Omega_o h_o$ for $\xi= 0.009$, when $\varepsilon= 10^{-2}$,
$\alpha_m=1$. The jet becomes super-SM slightly above the disc, meets the
Alfv\'en surface at $z_A \simeq 9.6 r_o$, where $r_o$ is the field line
anchoring radius on the disc, and finally stops when it meets the FM
critical surface. See FP95 for the expressions of these critical
velocities; in particular, $V = \vec{n}\cdot \vec{u}$ is not the jet
poloidal velocity.}
\label{fig:vit}
\end{figure}

The refocusing of the magnetic surfaces towards the jet axis stops because
the flow meets the fast-magnetosonic (FM) critical point (see
Fig. \ref{fig:vit}). The associated Mach number is defined as $t^2=  
V^2/V_{FM}^2$, where $V \simeq u_r$ at those altitudes (FP95). In the limit
of super-Alfv\'enic speeds ($g \simeq 1$), this Mach number can be written
as 
\be
t^2 \simeq \omega_A \left (\frac{2\lambda -3}{\lambda}\right)^{1/2}
\frac{r^2}{r^2_A} \sin^2\theta \, 
\ee
\noindent where $\theta$ is the jet opening angle (Fig. \ref{fig:jv4}). In
the other extreme limit ($g \simeq 0$), one has $t^2 \simeq
m^2\sin^2\theta$. Thus, as the jet refocuses towards the axis (but with
$r>r_A$), its opening angle grows and $t^2=1$ is unavoidable.   

The integration scheme stopped because no regularity condition was
imposed. A priori, such a critical point could (and should) be smoothly
crossed if, inside our parameter range, we could obtain two characteristic
behaviours at its vicinity. But we never obtained ``breeze''-like
solutions, all of them ending like in Fig. \ref{fig:vit}, in a way very
similar to Fendt et al. (\cite{fend}). Since we believe that our parameter
space is weakly affected by the self-similar ansatz used, it seems
to us doubtful that any physical trans-FM solution could be found. One
should bear in mind that, to this date, there is no semi-analytical jet
solution propagating to infinity that crosses all three critical points
(slow, Alfv\'en and fast). Providing a precise answer to whether such a
solution could exist is beyond the scope of the present paper. However, we
will give next section an argument against such a possibility, based on the
physics that leads to this critical situation.

Note that all jets displayed here achieve a super fast-magnetosonic
poloidal speed, namely an usual fast-magnetosonic Mach number $n^2=
u_p^2/V_{FM}^2 > 1$. The last critical point ($t^2=1$) would in principle
constrain another parameter of the MAES (e.g. the ejection index
$\xi$), thus leaving free (but severily bracketed) mostly one parameter,
the disc aspect ratio $\varepsilon$ ($\alpha_m$ is supposed to be provided
by calculations of MHD turbulence inside the disc).

Recollimation of magnetic surfaces is an old result: all BP82's solutions
displayed such a ``turning radius'', while PP92 found that such a
behaviour could be generic for cold jets (they did not use any self-similar
ansatz). Contopoulos \& Lovelace (\cite{cont:love}) however, using also a
self-similar ansatz, obtained jets with different asymptotic behaviours:
recollimating, ever-widening and oscillating. But these solutions were
obtained by varying the magnetic configuration parameter ($\beta$)
independently from the others ($\lambda$ and $\kappa$) and so, cannot
describe jets from Keplerian discs. These different behaviours reflect the
richness of the equations governing MHD jets.

Following PP92, let us define $\chi = m^2 r^2_A/r^2$ as a measure of the jet
widening. At the turning point ($B_r^t=0$), Bernoulli equation writes
\be
\chi^2_t \simeq \alpha_t^2 \kappa^2 \lambda^2 \ ,
\label{eq:chit}
\ee
\noindent which, in the limit $g_t \simeq 1$, provides the cubic  
\be 
\chi^3_t - (2\lambda -3)\kappa^2 \lambda^2(\chi_t - 1) \simeq 0 \ .
\ee
\noindent In that limit, one has $\chi = n^2/2 -1$ (PP92). Solving the
cubic and using this last expression, allows us to show that jets will
recollimate when they reach  
\be
n^2 > n^2_t = 2\kappa\lambda(2\lambda -3)^{1/2} - 3 \ .
\ee
\noindent In the other limiting case ($g_t \simeq 0$), Eq.(\ref{eq:chit})
provides directly the condition 
\be
n^2 > n^2_t = \kappa\lambda(\lambda -3)^{1/2} \ .
\ee
\noindent Both criteria, similar to those found by BP82 and PP92, are
indeed verified numerically for extreme cases (all solutions have
$n^2>1$). However, we used only Bernoulli equation to derive them: it 
is the jet transverse equilibrium that will tell us whether or not such a
possibility is indeed verified. Thus, we haven't provided yet a
satisfactory answer to why do jets recollimate. 

\subsection{Why do self-similar jets from Keplerian discs always
recollimate ?} 

\begin{figure}
%\picplace{4cm}
\psfig{figure=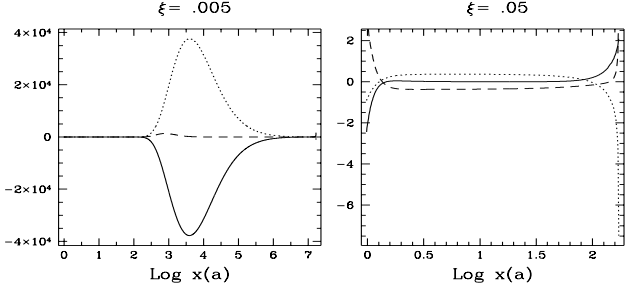,height=4truecm,width=8.8truecm}
\caption[ ]{Forces perpendicular to a given magnetic surface, along it,
in units of $B^2_p/\mu_or$, for $\alpha_m=1$, $\varepsilon=0.01$ and
$\xi=0.005$ (left), $\xi=0.05$ (right). The solid line is the sum of
gravitational, centrifugal and hoop-stress, the dotted line is the total
pressure gradient and the dashed line is the magnetic structure
response (namely, the term $(1-m^2)r/{\cal R}$). Confining forces are
negative. In the high-$\omega_A$ limit (left), it is the hoop-stress that
is responsible for recollimation (Log $x_A = 1.92$, Log $x_t= 3.6$). In the
low-$\omega_A$ limit (right), it is the pressure gradient associated to the
poloidal field (Log $x_A = 2.1$, Log $x_t= 2.2$).}  
\label{fig:jv5}
\end{figure}

In order to understand what forces the jet to bend towards the axis, one
can fruitfully look at another form of the jet transverse equilibrium
equation (\ref{eq:GS}), namely 
\begin{eqnarray}
(1-m^2) \frac{B^2_p}{\mu_o {\cal R}} & - & \nabla_{\perp}\left(P +
\frac{B^2}{2\mu_o}\right) - \rho \nabla_{\perp} \Phi_G \nonumber \\
& + &  (\rho\Omega^2r - \frac{B^2_{\phi}} 
{\mu_o r} )\nabla_{\perp}r = 0
\label{eq:fperp}
\end{eqnarray}
\noindent where $\nabla_{\perp}\equiv \nabla a \cdot \nabla /|\nabla a | $
provides the gradient of a quantity perpendicular to a magnetic surface
($\nabla_{\perp} Q <0$ for a quantity $Q$ decreasing with increasing magnetic 
flux) and ${\cal R}$, defined by
\be
\frac{1}{{\cal R}} \equiv \frac{\nabla a}{|\nabla a|} \cdot 
\frac{(\vec{B}_p\cdot\nabla)\vec{B}_p}{B^2_p} \ ,
\ee
\noindent is the local curvature radius of a particular magnetic surface
(Appl \& Camenzind \cite{appl:came93a}). When ${\cal R} >0$, the surface is
bent outwardly while for ${\cal R} <0$, it bends inwardly. The first term
in Eq.(\ref{eq:fperp}) describes the reaction to the other forces of both
magnetic tension due to the magnetic surface (with the sign of the
curvature radius) and inertia of matter flowing along it (hence with
opposite sign). The other forces are the total pressure gradient, gravity
(which acts to close the surfaces and deccelerate the flow, but whose
effect is already negligible at the Alfv\'en surface), and the centrifugal
outward effect competing with the inwards hoop-stress due to the toroidal
field (Fig. \ref{fig:jv5}). 

\begin{figure}
%\picplace{7cm}
\psfig{figure=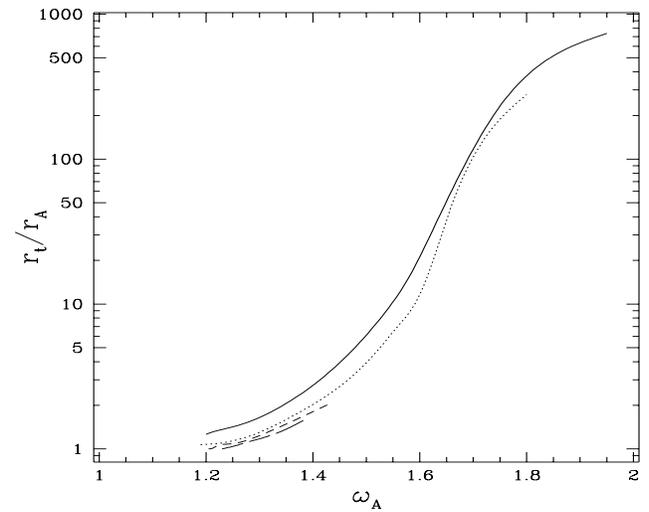,height=7truecm,width=8.8truecm}
\caption[ ]{Ratio of the maximum radius achieved $r_t$ to the Alfv\'en
radius, as a function of the fastness parameter $\omega_A$, for
$\alpha_m=1$ and $\varepsilon= 10^{-1}$ (solid line), $10^{-2}$ (dotted
line), $10^{-3}$ (short-dashed line), $7\ 10^{-4}$ (long-dashed
line). Regularity conditions at SM and Alfv\'en points fix the range of
allowed $\omega_A$. The jet is extremely sensitive to this parameter: jets
with $1< \omega_A < 1.5$ reach only a few times the Alfv\'en radius before
undergoing recollimation.}
\label{fig:para4}
\end{figure}

For current-carrying jets (high $\omega_A$, $g_A \la 1$), recollimation
occurs because of the constriction effect of the toroidal magnetic field
(as first discovered by BP82). For such highly super-Alfv\'enic solutions,
the jet transverse equilibrium depends mostly on the balance between this
force and the centrifugal one. Therefore, it will undergo recollimation
only if the jet widens enough and reaches the cylindrical radius
\be
\frac{r_t}{r_o} \simeq  \zeta_t^{-1/2} \kappa^{1/2} \lambda (2\lambda -
3)^{1/4} \ .  
\ee
\noindent where
\be
\zeta_t \equiv \left (\frac{\nabla_{\perp} \ln I}{\nabla_{\perp} \ln r}
\right )_t = \beta - 1 - \left ( \frac{\partial \ln B_{\phi}}{\partial \ln
x} \right )_t
\ee
\noindent describes how strongly the current varies across the magnetic
surfaces at ($z_t$ ,$r_t$). In the expression of the turning radius, the
other term scales as $\xi^{-3/4}$, implying that for smaller ejection indices,
the jet must open more in order to reach $r_t$. However, Fig. \ref{fig:psi}
shows that $r_t$ increases with $\xi$ much quicker than 
$\xi^{-3/4}$. Thus, $\zeta_t$ plays an important role and decreases very
rapidly with $\xi$: we found numerically that for $\varepsilon=0.01$,
$\alpha_m = 1$, we get $\zeta_t \simeq 1$, $10^{-1}$, $10^{-4}$ for $\xi =
0.015$, 0.01, 0.005 respectively. In fact, $\zeta_t$ is more generally a
function of the fastness parameter $\omega_A$ (which increases for
decreasing $\xi$): the faster the ``rotator'' and the bigger the maximum
radius reached (see Fig. \ref{fig:para4}). The centrifugal picture is
therefore quite appealing here, but one should not forget that $\omega_A$
is a measure of $I_A$. 

For current-free jets (low $\omega_A$, $g_A \ll 1$), $r_t \ga r_A$ and the
flow is only slightly super-Alfv\'enic. Here, the ratio of the toroidal
magnetic force to the centrifugal one writes  
\be
\frac{B_{\phi}^2}{\mu_o \rho \Omega^2 r^2} \frac{\nabla_{\perp} \ln
I}{\nabla_{\perp} \ln r} \simeq \zeta_t g_t^2 
\ee
\noindent with $\zeta_t$ of order unity but $g_t \ll 1$. Thus, the force
responsible for recollimation can only be the magnetic pressure gradient
associated with the poloidal field. Indeed, the jet transverse equilibrium
(\ref{eq:fperp}) implies that this force is negative, hence trying to close
the jet. If both centrifugal force and magnetic force associated to the
toroidal field cannot overcome it, the jet cannot avoid a recollimation
(Fig. \ref{fig:jv5}). If $\omega_A >1$ is a necessary condition to obtain
trans-Alfv\'enic jets, it is obviously not sufficient: the ``rotator'' must
be fast enough to allow the propagation of jets much farther away from the
Alfv\'en surface. Since this result appears at the Alfv\'en surface
vicinity, we believe that it is real, namely independent of our
self-similar modelling. 

\begin{figure*}
%\picplace{10.3cm}
\psfig{figure=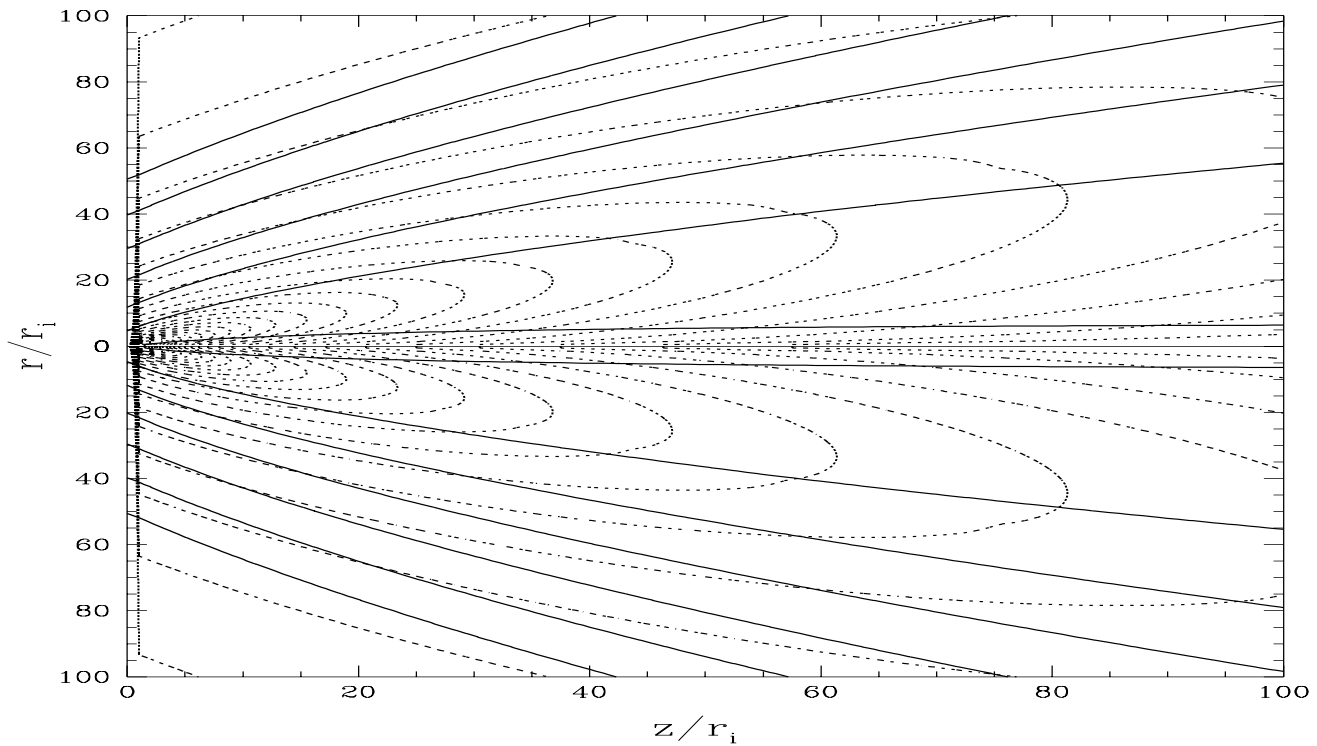,height=10.3truecm,width=18truecm}
\caption[ ]{Isocontours of poloidal current density ($\vec J_p$, dotted
lines) and poloidal magnetic field lines ($\vec B_p$, solid lines) for a
cold jet with $\xi= 0.01$, launched from a disc of $\varepsilon= 10^{-2}$,
$\alpha_m=1$. The current circuit displays a butterfly-like shape,
characteristic of tenuous ejections ($\xi < 1/2$). The current flows down
the jet axis, enters the disc at its inner edge $r_i$ and returns along the
jet itself. Force-free solutions would have $\vec J_p \parallel \vec B_p$,
which is not quite achieved here, even for smaller values of $\xi$. The jet
carries its own current, building up a global electric circuit as it
propagates through the interstellar medium.}
\label{fig:circuit}
\end{figure*}

We can now try to understand the fate of cold jets from Keplerian discs,
once they start to recollimate. At the turning radius, matter has almost
reached its maximum poloidal velocity, with $n^2 \gg 1$ (for high
$\omega_A$ jets). There is therefore no way to slow down the poloidal
motion ($F_{\parallel}$ remains too small) and, despite the decrease in the
jet radius (leading to a decrease in $m^2= \rho_A/\rho$), the poloidal
velocity remains roughly constant. Matter carries away the field with it,
strongly decreasing the jet pitch angle. Thus, the total current $I$ (or
equivalently $B_{\phi}$) goes to zero. Eventually, the dominant
magnetic pressure effect in Eq.(\ref{eq:fperp}) becomes the one due to the
poloidal field. Because of recollimation, the pressure gradient associated
with the poloidal field changes its sign and pinches the jet too (some sort
of depression). As a result, the jet is forced to bend towards the axis,
with a curvature radius ${\cal R}$ becoming infinitely negative. This
behaviour was also obtained by Contopoulos \& Lovelace (\cite{cont:love})
and lead these authors to propose that the assumption of steady-state
should break down at these distances. More generally, any steady-state jet
with zero poloidal current would be asymptotically parabolic (Heyvaerts \&
Norman \cite{heyv:norm}). Such a situation is very far from the actual
recollimating regime. Thus, it seems reasonable to guess that stationarity
breaks down at the jet ``end'' (where $t^2= 1$ is met but not crossed) and
that a shock is formed there.

It is interesting to note that, althought they used the same
self-similarity ansatz as we did, Contopoulos \& Lovelace
(\cite{cont:love}) found also non-recollimating (for $\beta < 1$) and
oscillating (for $\beta>1$) solutions. But, as already said, those were
obtained with parameters that do not correspond to disc-driven jets and
achieved regimes with $n^2<1$, thereby allowing a different asymptotic
state. On the same line of thought, any neglect of the poloidal field (like
in Contopoulos \cite{cont95}) would miss the magnetic depression effect due
to recollimation, hence modifying also the jet asymptotic equilibrium (and
possibly allowing trans-FM solutions by fine-tunning a parameter).

\subsection{Why do self-similar jets widen so much ?}

For current-carrying jets (Fig. \ref{fig:circuit}), it has been shown that
jets recollimate through the constriction action of the magnetic
field. This is possible because the jet radius keeps on opening, allowing a
huge acceleration efficiency as well as a decrease of the matter rotation
rate. Why do we obtain such a systematic behaviour ?

To address this question, it is worthwhile to look at the Grad-Shafranov
equation (\ref{eq:GS}), written in the following form 
\be
(1-m^2) J_{\phi} = J_{\lambda} + J_{\kappa} \ .
\ee
\noindent It expresses that the toroidal current is generated by two main
sources, namely 
\be
J_{\lambda}= \rho r \left \{ \frac{\dd {\cal E}}{\dd a} + (1-g)\Omega_*r^2 
\frac{\dd \Omega_*}{\dd a} + g\Omega_*\frac{\dd \Omega_*r^2_A}{\dd a }
\right \}
\ee
\noindent that depends mostly on how the magnetic lever arm behaves from
one magnetic surface to another, and
\be
J_{\kappa}= m^2 \frac{\nabla a}{\mu_o r}\cdot \nabla \ln \rho \ + \
r\frac{B^2_{\phi} - m^2B^2_p}{2\mu_o}\frac{\dd \ln \rho_A}{\dd a}
\ee
\noindent strongly dependent on the mass load at each surface. It is clear 
that the behaviour of global quantities has a tremendous importance on the
jet equilibrium, thus casting doubts on approaches that are not trully 
2-D. In the self-similar description of disc-driven jets and within our
parameter range, one always achieves a turning point ($B_r^t =0$) with
$\sigma_t \ll 1$. Since the toroidal current at this point, 
\be
J^t_{\phi} \simeq - \frac{1}{\mu_o} \frac{\partial B_z}{\partial r} \propto
\beta(2 - \beta)
\ee
\noindent is positive, recollimation is allowed only if the currents
$J^t_{\kappa}$ or $J^t_{\lambda}$ are negative. At this turning point, one
source writes
\begin{eqnarray}
J^t_{\kappa}  &\simeq &\frac{B^2_p r}{2 \mu_o B_o r_o^2} \left \{
\frac{B^2_{\phi}}{B^2_p}\left ( \frac{\dd \ln \rho_o}{\dd \ln r_o} + 
\frac{\dd \ln \rho_A/\rho_o}{\dd \ln r_o}\right ) \right . \nonumber \\
&& \hspace{1.3cm} + \ \left . m^2 \left ( \frac{\dd \ln \rho_o}{\dd \ln r_o}
- \frac{\dd \ln \rho_A/\rho_o}{\dd \ln r_o} \right ) \right \} \ ,
\end{eqnarray}
\noindent for all solutions. On the contrary, the other current source
strongly depends on $g$ (thus on $\omega_A$): for $g_t \simeq 1$ it writes 
\be
J^t_{\lambda} \simeq \frac{\rho \Omega^2_o r}{2B_o} \left \{
3 + 2\lambda \left ( \frac{\dd \ln \lambda}{\dd \ln r_o} - 1 \right) \right \}
\ee
\noindent and for $g_t \simeq 0$ 
\be
J^t_{\lambda} \simeq \frac{\rho \Omega^2_o r}{2B_o} \left \{ 3 -3\lambda
\right \} \ . 
\ee

Self-similar solutions necessarily imply that ratios like $r_A/r_o$ and
$\rho_A/\rho_o$ are constant through all the jet. Their logarithmic
derivatives are therefore zero and both current sources are indeed 
negative at the turning point ($\dd \ln \rho_o/\dd \ln r_o = 2\beta - 3=
-3/2 + \xi <0$, FP93a). Thus, our systematic behaviour arises from both our
MAES parameter space and the fulfilment of this condition. On the contrary,
any jet with $g_t \sim 1$ (high $\omega_A$) but satisfying
\begin{eqnarray}
\frac{\dd \ln r_A/r_o}{\dd \ln r_o} & > & \frac{1}{2} \nonumber \\ 
\frac{\dd \ln \rho_A/\rho_o}{\dd \ln r_o} & < & \frac{\dd \ln\rho_o} {\dd
\ln r_o}  
\label{eq:cond}
\end{eqnarray}
\noindent would never allow $J^t_{\phi}>0$ and therefore, would not undergo
recollimation. This sufficient condition expresses that high-$\omega_A$
jets would not widen so much as they do here, the magnetic structure
keeping stored a significant amount of the power provided by the
disc. Since our parameter space is only slightly dependent on our
self-similar modelling, we claim that it describes realistic boundary
conditions for any given magnetic surface. Therefore, we expect that any
model of a magnetically-driven jet would also undergo recollimation if 
conditions (\ref{eq:cond}) are not met. ``Over-widening'' and then
recollimation would be general features, independent on the analytical
model used, of cold jets from discs described by constant parameters. This
is consistent with PP92, who did not use any self-similar assumption, but
found recollimating solutions by making the approximation $\omega_A$ (their
$\lambda$ parameter) constant through the jet. Sakurai (\cite{saku87})
performed full numerical simulations of jets from accretion discs that
crossed the three usual critical points (measured with the poloidal
velocity), but without taking self-consistently into account the
disc. Because of the limited computational domain, it is difficult to 
figure out whether or not his solutions display a recollimation (see the
vertical scale involved for small $\xi$ in Fig.\ref{fig:psi}). However, 
his initial split-monopole geometry for the magnetic configuration
forbids the development of a jet with a constant $\xi$ (see his Fig.3). To
determine if the final configuration is realistic would require a full 2-D
treatment of the disc-jet interrelations.

For current-free or low-$\omega_A$ jets ($g_t \sim 0$), non-recolli\-mating
solutions could in principle be obtained if 
\be
\frac{\dd \ln \rho_A/\rho_o}{\dd \ln r_o} < \frac{\dd \ln\rho_o} {\dd \ln
r_o} - 3 \frac{\lambda -1}{\lambda - 3} 
\ee
\noindent is verified. This necessary condition is much more stringent that
in the case of high-$\omega_A$ jets, and we have strong doubts that it
could ever be met. As already mentionned however, we expect that the
behaviour of these kind of jets is mostly independent of our modelling. 

It is noteworthy that differential rotation is of so great importance for
both jet formation and collimation. Indeed, it has already been found that
without the counter current due to the disc differential rotation, no
magnetically-driven jet could be possible. Here, we find that in the
absence of differential rotation ($\dd \Omega_*/\dd a = 0$), one of the two
sources of toroidal current remains positive for any $g_t$, namely 
\be
J^t_{\lambda} \simeq \frac{\rho \Omega^2_o r}{2B_o} \left \{
3 + g_t\lambda \left ( 4 + 2\frac{\dd \ln \lambda}{\dd \ln r_o}\right)
\right\}  \ .
\ee
\noindent Thus, differential rotation of magnetic surfaces has a tremendous
importance in recollimating outflows, or more generally, in their
asymptotic behaviour. This is a very important effect and shows that jets
from a rigidly rotating object (e.g., a star) will certainly have a
different asymptotic behaviour than jets from accretion discs. In
particular, they can display non-recollimating or oscillating behaviours
(Tsinganos \& Trussoni \cite{tsin:trus}, Sauty \& Tsinganos
\cite{saut:tsin}). 

MAES are expected to be settled in the innermost part of a larger accretion
disc, where both the energy reservoir and magnetic field strength are
bigger. Such a picture implies that a viscous-like transport of angular
momentum acts in the outer disc until magnetic braking due to the jet
overcomes it in the inner regions. One would then obtain a transition, as
$r_o$ decreases, from $\xi= 0$ to $\xi \la 0.01$ with the corresponding
changes in $r_A/r_o$ (from infinity to a finite value) and $\rho_A/\rho_o$
(from zero to a finite value). Thus, the conditions (\ref{eq:cond}) could
be naturally fulfilled in a self-consistent picture taking into account
realistic boundary conditions. Needless to say that only 2-D numerical
calculations could achieve it.

\section{Summary and conclusion}

In this paper, we have covered all dynamical processes at work in a
stationary, weakly dissipative MAES of bipolar topology (see Appendix
\ref{Ap:quadru} for quadrupolar), by constructing continuous solutions 
from the accretion disc to super-Alfv\'enic jets. We summarize below our
findings and discuss their implications.  

(1) We have demonstrated that the usual parameters for magnetically-driven
(cold) jets, namely the magnetic configuration, lever arm and mass load, are
intrinsically related to the disc ejection efficiency $\xi$. Thus,
realistic jets would be described by a distribution of $\xi$ with the
magnetic surfaces (or their anchoring radius). 

(2) We showed that this parameter lies in a very narrow range for
cold jets, namely $0.004 \la \xi \la 0.08$. A minimun ejection efficiency is
required for the disc to find a quasi-MHS equilibrium. A crude modelling of
this sensitive equilibrium leads most probably to unstable regimes, the disc
being too much pinched by the Lorentz force. On the other hand, a maximum
mass load arises from the constraint of accelerating matter up to
super-Alfv\'enic speeds. The threshold for the minimum ejection index is
raised for decreasing disc aspect ratio $\varepsilon$ and turbulence
parameter $\alpha_m$. Both constraints impose that these parameters must
verify $0.0005 < \varepsilon \le 0.1$ and $0.1 < \alpha_m \le 1$. 

(3) As a result, the current must enter the disc at its inner edge, flow back
in the jet and close by flowing down along the axis. This has two important
consequences. First, although the disc can afford larger ejection efficiencies
(Ferreira \cite{these}), they would be non-steady, the flow not being able to
reach the Alfv\'en surface. Second, this implies a strong influence on
what's going on at the axis, thus suggesting a coupling with the inner
central object magnetosphere. This work is under progress. 

(4) The existence of a minimum ejection index has strong implications on
jet energetics. Indeed, it forbids to construct jet models where the mass
load is arbitrarily small (and thus, an asymptotic velocity arbitrarily
high). Nevertheless, in the range allowed for $\xi$, the ratio of the total
ejection rate (in one jet) to the accretion rate, $f \equiv {\dot
M}_{je}/{\dot M}_{ae}$, would vary between $10^{-3}$ and $10^{-1}$,
depending on both $\xi$ and the radial extension $r_e/r_i$ of the
magnetized disc (Ferreira \cite{F96}). This is in complete agreement with
recent estimates made by Hartigan et al. (\cite{harti95}), who find a
typical value of $10^{-2}$ for jets from young stars. 

The jet magnetization parameter (Michel \cite{mich}, Camenzind \cite{came87}),
\be
\sigma_* \equiv \frac{\Omega^2_*B_o r_o^2}{\eta c^3} = 6.8 \ \left(
\frac{\sigma_{SM}}{100} \right ) \left (\frac{-B_o}{2B_{\phi,SM}} \right)
\left(\frac{r_o}{3r_g}\right)^{-3/2} 
\ee
\noindent shows that relativistic jets could be possible around compact
objects. If Compton drag can be avoided (Phinney \cite{phin87}), the mean
bulk Lorentz factor $\bar \gamma$ that such jets could achieve, 
\be
\bar \gamma = 1 + \frac{\eta_{lib}}{24 f} (1 + \bar \sigma_{\infty})^{-1}
\left(\frac{r_i}{3r_g}\right)^{-1}
\ee
\noindent strongly depends on how much power the magnetic structure
keeps stored in its asymptotic regime (see FP95 for the definition of
$\eta_{lib}$). As an example, for $\bar \sigma_{\infty} = 1$ (final kinetic
energy flux comparable to the Poynting flux, Li et al. \cite{li92}), one
gets $\bar \gamma$ between 1 ($\xi= 0.07$, $r_e=100 r_i$) and 8 ($\xi =
0.004$, $r_e=2 r_i$), with typical values lying between 1 and 4. Thus
moderate mean bulk Lorentz factors are likely to be achieved by cold jets
from compact objects.

In the context of YSOs, the mean jet velocity writes 
\be
\bar u_{\infty} = \Omega_i r_i (1 + \bar \sigma_{\infty})^{-1/2} \left(
\frac{\eta_{lib}}{2f}\right)^{1/2}  \ ,
\ee
\noindent where $\Omega_i r_i$ is the angular velocity at the inner disc
radius. For $\bar \sigma_{\infty} = 1$, jets can reach a velocity between 1
($\xi= 0.07$, $r_e=100 r_i$) and 10 ($\xi = 0.004$, $r_e=2 r_i$) times this
velocity, with typical factors between 2 and 6. For structures settled at
$r_i \simeq 30 r_{\sun}$ (10 stellar radii for a typical T-Tauri star),
this provides jets with a mean velocity between 100 and 500 kms$^{-1}$, in
agreement with observations.

(5) All the above results are general to magnetically-driven jets from
Keplerian accretion discs. We constructed global solutions, from the
accretion disc to a super-Alfv\'enic jet, using a self-similar ansatz. It
has been shown that the jet asymptotic behaviour strongly depends on the
fastness parameter $\omega_A$, which describes how fast is the magnetic
``rotator'' and is a measure of the amount of current still available at
the Alfv\'en surface. This parameter must be bigger than (but of the order
of) unity and is very sensitive to the physical conditions inside the
disc. All our solutions display the same behaviour: the jet widens until a
very strong acceleration efficiency is achieved, all the available power
being eventually transferred into kinetic power; the centrifugal force
decreases so much that the Lorentz force pinches the jet, making it to
recollimate. The bigger $\omega_A$ and the larger the maximum radius
reached by the jet. Inside our parameter range (which is weakly affected by
our self-similar assumption), this behaviour stems from having a constant
ejection index $\xi$. Therefore, we expect that non self-similar jets from
Keplerian discs described with constant parameters would also display
recollimation, unless realistic 2-D boundary conditions are taken into
account. Such boundaries should concern the transition from the outer
viscous disc to the inner magnetized one, as well as a possible interaction
with the central object.

(6) If magnetized discs are driving jets over a wide range of radii, then
they can probably be described by an almost constant ejection index $\xi$
(and self-similar solutions are not too bad an approximation). In those
circumstances, one would expect that these jets undergo a recollimation and
then a shock, making the whole structure unsteady. Whether or not such a
shock is terminal or a ``magnetic focal point'' (Gomez de Castro \& Pudritz
\cite{gome:pudr}, Ouyed \& Pudritz \cite{ouye:pudr}) remains to be
carefully worked out. Besides the shock signature, such a structure could
be detected through its apparent lack of radiation coming from the
magnetized disc itself (FP95). However, one has to bear in mind that the
fate of jets could be strongly modified by the external medium (Appl \&
Camenzind \cite{appl:came93a}). Indeed, an external confinement could
forbid a natural ``over-widening'', thus enforcing a redistribution of the
current density inside the jet. By this way, asymptotic solutions with
$\nabla_{\parallel} I = 0$ but with non-vanishing current $I$ (in contrast
with what is obtained here) could perhaps be achieved. Such an hypothesis
should deserve further investigation.

(7) Finally, a major challenge remains the question of the source for the
required magnetic diffusivity. Indeed, Heyvaerts et al. (\cite{heyv96})
showed that in a disc braked by viscous stresses, any instability (possibly
magnetic), triggering a turbulence with an injection scale of order the
disc thickness, would provide ${\cal R}_m \sim 1$. Such a situation is
therefore incompatible with magnetically-driven jets. As a consequence,
these powerful jets require physical conditions that were not yet
investigated in discs like those, for example, met at the interaction with
the central object magnetosphere. On the other hand, thermally-driven jets
(but magnetically confined) should be viewed as a possible alternative. The
jet power could be reduced down to a level comparable to the disc luminosity
(see Eq.(\ref{eq:Pj}), with $\Lambda \sim 1$). Such a picture is appealing
because it allows jets along with radiating discs and offers a smooth
transition between ``viscous-like'' discs and MAES (Ferreira, in
preparation).

\begin{acknowledgements} 
I would like to thank the referee, Kanaris Tsinganos, for his helpful
comments, as well as Max Camenzind, Jean Heyvaerts, Stefan Appl, Ramon
Khanna and Guy Pelletier for enthousiastic and stimulating
discussions. This work was supported by the Deutsche Forschungsgemeinschaft
(SFB 328). 
\end{acknowledgements}

\appendix

\section{Appendix: Quadrupolar topology and the production of jets}
\label{Ap:quadru}

In this appendix, we briefly investigate the structure of a Keplerian
accretion disc, thread by a large scale magnetic field of quadrupolar
topology. Such a topology is described by an even toroidal field and a
magnetic flux $a(r,z)$ (see Eq.(\ref{eq:topo})), being an odd function of
$z$. The magnetic field inside the disc follows the plasma, with
$B_{z,o}=0$, $B_{r,o} <0$ and $B_{\phi,o} >0$ (subscripts ``o'' refer to
quantities at the disc midplane). Since both plasma angular velocity and
poloidal magnetic field increase towards the center (we don't take into
account a possible boundary layer between the disc and the central
object), the disc can be divided into two distinct regions (see Lovelace et
al. \cite{love87}, Khanna \& Camenzind \cite{khan:came92}). The inner
region is characterized by a positive vertical current density at the disc
midplane ($J_{z,o} >0$), but negative at its surface ($J_z^+ <0$). The outer
region displays the opposite behaviour, with $J_{z,o} <0$ and $J_z^+ >0$
(the transition between the two is of course at the radius where $J_{z,o} =
0$).
 
One would like to build up a Keplerian disc (that is, in quasi-MHS
equilibrium in both radial and vertical directions), accreting towards the
center ($ u_{r,o} < 0$) and giving rise to a magnetically propelled jet. This
last demand requires the following (necessary) conditions: (a) open field
lines ($B_z^+$ and $B_r^+$ both positive), (b) a positive MHD Poynting flux
(for $z >0$, namely $B_{\phi}^+ <0$) and (c) a MHD acceleration
($\nabla_{\parallel} I >0$, see Eq.(\ref{eq:force})). 

The inner region is unable to steadily produce MHD jets. This arises from the
topology itself, which provides $F_{\phi}^+ <0$ (therefore $F_{\parallel}
<0$): the magnetic structure would slow down any plasma that could
have been (thermally) ejected out from the underlying disc. 

On the contrary, the outer region can in principle produce jets. Let us
then examine its structure. Since the Lorentz force is accelerating the
matter at the disc midplane, accretion is achieved only if the viscous
torque is dominant ($\Lambda < 1$). Thus, the accretion velocity is
\be
u_o = - u_{r,o} \simeq 2 (1 - \Lambda) \frac{\nu_{v,o}}{r} \ ,
\ee
\noindent where $\nu_{v,o}$ is the anomalous viscosity (measured at the
disc midplane). Both radial and vertical components of the Lorentz force
are positive, thus counter-acting gravity inside the disc. If the disc is
in quasi-MHS vertical equilibrium, then the deviation from Keplerian
rotation law is of the order $\varepsilon^2$. At the disc surface, the
vertical Lorentz force changes its sign, keeping nevertheless 
$F_{\parallel}>0$ if $J^+_z > J^+_r B^+_z/B^+_r$ is satisfied (condition
(c)), with $J^+_r >0$. Condition (a) is fulfilled at the disc surface only
if the vertical scale of variation of the magnetic flux is of the order of
the disc scale height. Steady-state diffusion of the poloidal field
(Eq.(\ref{eq:diff})) requires then a ``poloidal'' magnetic diffusivity such
that $\nu_{m,o} \simeq \varepsilon u_o h$. The toroidal field at the disc
surface can be written as $B^+_{\phi} = B_{\phi,o} + B^{\nabla
\Omega}_{\phi}$, where $B^{\nabla \Omega}_{\phi}$ is provided by the disc
differential rotation (Eq.(\ref{eq:ind})). Since $B^+_{\phi}$ must be
negative (condition (b)),  
\be
\frac{B_{\phi o}}{- B_{ro}} \la \frac{3}{2} \frac{\Omega_o r}{u_o} 
\frac{\nu_{m,o}}{\nu_{m,o}'}
\ee
\noindent where $\nu_{m,o}'$ is the ``toroidal'' diffusivity (see
Sect. 2.1). With these estimates, one obtains that with the following
restrictions, namely
\begin{eqnarray}
&  & \frac{B^2_{r,o}}{\mu_o P_o} \simeq \Lambda^2 \left (\frac{\nu_{v,o}}
{\Omega_o h^2} \right)^2 \nonumber \\
&  & \frac{B^2_{\phi,o}}{\mu_o P_o} < 1 \nonumber \\
&  & \frac{\nu_{m,o}}{\nu_{m,o}'} \la \frac{4}{3} (1 - \Lambda)
\frac{\varepsilon^2} {\Lambda} \\
&  & \frac{\nu_{m,o}}{\nu_{v,o}} \simeq 2 (1 - \Lambda) \varepsilon^2
\nonumber 
\end{eqnarray}
\noindent where $\Lambda < 1$ and the plasma pressure $P_o = \rho_o
\Omega_o^2 h^2$ (quasi-MHS vertical equilibrium), quadrupolar topologies
fulfill the disc requirements as well as the jet conditions (a) and (b)
(condition (c) requires a full calculation of the vertical structure). Note
that above the disc surface, the current system is similar to the one of a
bipolar magnetic topology: the magnetic energy that feeds the jet arises
from the disc plasma itself, extracted where $F_{\phi}<0$ (in a layer
around $J_z = 0$). Since the torque there cannot be as strong as in the
bipolar case, one gets the general result that jets from quadrupolar
topologies would be weaker.

However, this configuration demands very peculiar conditions on the disc
turbulence. Indeed, a situation giving rise to $\Lambda \la 1$ requires
$\nu_m \sim (1-\Lambda) \varepsilon^2\nu_m'$ with $\nu_m' \sim \nu_v$,
whereas $\Lambda \sim \varepsilon$ is achieved for $\nu_m \sim \varepsilon
\nu_m' \sim \varepsilon^2 \nu_v$, and $\Lambda \sim \varepsilon^2$ for
$\nu_m \sim \nu_m' \sim \varepsilon^2 \nu_v$. Such a situation is against
our current understanding of turbulence, where all transport coefficients
should achieve a comparable level (Pouquet et
al. \cite{pouq76}). Therefore, it is dubious that such a topology could 
produce MHD jets. It could nevertheless play a role in a quadrupolar
``cored apple'' circulation around protostellar sources (Henriksen \&
Valls-Gabaud \cite{henr:vall}, Fiege \& Henriksen \cite{fieg:henr}).

\section{Appendix: Toroidal field at the disc surface}
\label{Ap:Bphi}

We start with Eq.(\ref{eq:Bphi}), by making a second order Taylor expansion
of all the quantities and neglecting the advection term, and we obtain 
\be
\frac{\eta_m' J_r}{\eta_o'J_o} \simeq 1 - \Gamma \left( \frac{x^2}{2}
+ 2\frac{\Omega_o -\Omega}{3\Omega_o {\cal R}_m \varepsilon^2} \right)
\ee
\noindent where $\Gamma = 3\nu_o/ \alpha_m^2\nu_o'$ is a measure of the
degree of anisotropy of the MHD turbulence inside the disc. The disc radial
equilibrium provides the angular velocity $\Omega = \Omega_k (1 + \omega)$,
where the deviation from Keplerian rotation law comes mainly from the
radial magnetic tension,
\be
\omega \simeq - \frac{\mu}{2}{\cal R}_m \varepsilon^2 v
\ee
\noindent with $v \equiv \rho_o/\rho \simeq 1 + v_o''x^2/2$. The above
expression shows that, as the density decreases vertically, the magnetic
effect increases and the plasma rotates with a lower rate. This gives rise
to an enhanced accretion velocity above the disc midplane (and a
corresponding source of toroidal current, see Figs. 4, 5 and 8 in
FP95). The disc vertical equilibrium provides the density profile, namely
\be
v_o'' \simeq 1 \;+\; \mu q^2 \;+\; \mu {\cal R}_m^2\varepsilon^2 
\ee
\noindent where the first term is the tidal force, the second one is the
magnetic pressure due to shear ($B_{\phi}$) and the last one is the
magnetic pressure due to curvature ($B_r$). These three effects are
comparable (see Fig. \ref{fig:para3}). Using that the ``toroidal''
magnetic resistivity decreases vertically as $\eta_m' = \eta_o'(1 -
d_ox^2/2)$, we get 
\be
\frac{J_r}{J_o} \simeq 1 - \delta \frac{x^2}{2} \mbox{ \hspace{.5cm} with
\hspace{.5cm}} \delta = \Gamma\left( 1 + \mu\frac{v_o''}{3}\right) - d_o 
\ee
\noindent and therefore, $B_{\phi} = B_{\phi}^{J_o} + B_{\phi}^{\nabla 
\Omega}$, where $B_{\phi}^{J_o} = - q B_o x$ and $B_{\phi}^{\nabla \Omega}
= q B_o \delta x^3/6$. The toroidal field inside the disc is then mainly
measured by $B_{\phi}= -qB_o f(x)$, with $f(x)= x(1 - \delta x^2/6)$. Since
the magnetic diffusivity decreases on a disc scale height ($d_o$ of 
order unity), the only way to allow a decrease of $J_r$ on such a scale
($\delta>0$ and of order unity) is to require $\Gamma \simeq 1$. This is the
key factor for ejection (Ferreira \& Pelletier \cite{FP93b}, FP95). This,
in turn, implies $f_+ = f(1) \simeq 0.5$ to 1. Such an important result is
independent of the diffusivity used, as long as it decreases on a disc
scale height.

\section{Appendix: Minimum ejection index}
\label{Ap:min}

The disc is not in a perfect MHS equilibrium, but there is a slight motion
towards the disc midplane. From mass conservation, we obtain that this tiny
motion is
\be
u_z \simeq \varepsilon u_o (\xi - 1) x \ .
\ee
\noindent Thus, this velocity increases as $\xi$ diminishes, describing the
basic fact that the plasma pressure gradient is less effective in
sustaining the disc against both tidal and magnetic compression. This then
implies that $v_o''$ decreases (see Appendix \ref{Ap:Bphi}), which leads to
a decrease of  $\delta$, hence an increase of $B_{\phi}$ at the disc
surface ($f_+$ increases). Now, a look at Eq.(\ref{eq:vertmu}) shows that a
vertical equilibrium will not be achieved for $f_+$ too high (how high
being obtained only by numerical means).

Equation (\ref{eq:Bphi}), taking into account the advection term, can be
written as 
\be
\frac{\eta_m' J_r}{\eta_o' J_o} \simeq 1 - \Gamma \left ( 1 +
\mu\frac{v_o''}{3} + \frac{2}{3}(\xi - 1)\alpha_m^2{\cal R}_m\varepsilon^2 
\right )\frac{x^2}{2} \ . 
\ee
\noindent This expression shows that inside the disc, where this expansion
is valid, the effect of advection is to decrease the radial current
density (although this contribution, of order $\alpha_m^2\varepsilon$, is
negligible). On the contrary, above the disc when $u_z >0$, this term
changes its sign and contributes to maintain $J_r >0$. Eventually, as one
goes from the resistive disc to the ideal MHD jet, this term will completely
balance the effect of differential rotation. What the above expression
shows, is that its influence will start at lower altitudes for higher
$\alpha_m$ and $\varepsilon$. Thus, to the highest values of both
$\alpha_m$ and  $\varepsilon$ corresponds the highest value of $f_+$,
henceforth the smallest ejection index.

\end{document}